\documentclass[10pt,letterpaper]{article}

\usepackage[margin=0.85in,top=0.8in,bottom=0.9in]{geometry}

\usepackage{times}
\usepackage[T1]{fontenc}
\usepackage[utf8]{inputenc}
\usepackage{microtype}
\usepackage{graphicx}
\usepackage{booktabs}
\usepackage{amsmath,amssymb}
\usepackage{multirow}
\usepackage{array}
\usepackage{caption}
\usepackage{subcaption}
\usepackage{listings}
\usepackage{xcolor}
\usepackage[colorlinks=true,allcolors=blue,pdftitle={CausalOdyssey:
Multi-Agent Local LLM Reasoning with Per-Patient Causal Structure for Rare
Disease Retrieval},pdfauthor={Xueer Chen}]{hyperref}

\usepackage{enumitem}
\setlist{nosep,leftmargin=1.2em}

\usepackage{titlesec}
\titlespacing{\section}{0pt}{1.0em}{0.5em}
\titlespacing{\subsection}{0pt}{0.7em}{0.3em}

\title{\vspace{-1.0em}\textbf{CausalOdyssey: Multi-Agent Local LLM Reasoning with Per-Patient Causal Structure for Rare Disease Retrieval}}

\author{Xueer Chen\\[0.2em]
{\normalsize\href{mailto:xueer.chen.human@gmail.com}{\texttt{xueer.chen.human@gmail.com}}}\thanks{This work was conducted independently, outside the author's employment, using only public data and open-weight models. The views expressed are the author's own and do not represent any employer.}}
\date{}

\begin{document}
\maketitle
\vspace{-2.0em}

\begin{abstract}
\small
Patients with rare diseases often wait years for a diagnosis. Existing HPO-based retrieval tools narrow the differential to a shortlist, but a clinician reading that shortlist has no signal for which candidate to trust or what to test next. We present \textbf{CausalOdyssey}, a local four-agent pipeline on Gemma~4~26B built over a classical IC-weighted gene-rescue backbone: a \emph{Diagnostician} reranks the slate via chain-of-thought over per-patient FCI causal edges and Reactome mechanism context; a \emph{Verifier} computes deterministic evidence against the top-1 (missing pathognomonic phenotypes, pathway chain closure) and has Gemma issue a confirm/flag/refute verdict over it; an \emph{Interrogator} selects one follow-up HPO by split-variance scoring; a \emph{Narrator} emits citation-validated clinical prose. Only the Diagnostician changes the ranking; the other three agents annotate it. On RareBench $n{=}1{,}122$, reranking a top-30 slate lifts $R@1$ by +12.57\,pp over our structural baseline ([+10.07,\,+14.97]; $p{=}8.3{\times}10^{-24}$). A no-LLM variant of that baseline, without its inverse disease-count weight, reaches a similar pooled $R@1$ ($36.3\%$ vs $35.3\%$): the LLM rerank is far better on metabolic cases (RAMEDIS, +16.0\,pp) and far worse on case-report corpora (LIRICAL, MME), so its value is corpus-dependent. We also propose a masked-phenotype protocol for evaluating follow-up suggestions without an annotation-based oracle; under it, the Interrogator asks for a patient's hidden phenotypes no more often than simple baselines ($0.6\%$ vs $1.2\%$), although a perfect follow-up would add $11$\,pp at top-1. The Verifier's verdict separates correct from incorrect top-1s ($51.4\%$ vs $12.1\%$ top-1 accuracy under confirm vs refute). For 30\% of patients the gold diagnosis is outside our top-30, though LIRICAL retrieves over half of these. All inference is local; no patient data leaves the machine. Demo code and a live demo are public.
\end{abstract}

\section{Introduction}

Rare disease affects approximately 300~million people worldwide, and the time to a confirmed diagnosis commonly extends over several years~\cite{rarebench}. The odyssey is driven less by disease prevalence than by the gap between phenotype observation at the clinic and retrieval of the handful of candidate diagnoses a geneticist would consider. The Human Phenotype Ontology (HPO)~\cite{hpo} bridges part of that gap: clinicians encode findings as HPO terms, and retrieval systems such as Phen2Gene~\cite{phen2gene}, LIRICAL~\cite{lirical}, and PubCaseFinder~\cite{pubcasefinder} return a ranked shortlist of candidate genes or diseases via phenotype-similarity or likelihood-ratio scoring. Under the best conditions these tools place the correct diagnosis in the top ten for roughly half of pediatric metabolic cases. But the patients who most need a diagnosis --- those in rural or low-connectivity regions --- are the same patients for whom cloud-API tools are unreachable: institutional review rules often forbid sending identifiable phenotype tuples off-premise, and rural clinics cannot rely on bandwidth that supports LLM inference even when permitted. Consider an illustrative composite (not a real case): a father in Apur\'{\i}mac, Peru describes his five-year-old daughter in Quechua; a community health worker translates to Spanish; a pediatrician writes six structured HPO codes. The system that reads those codes must run on hardware the clinic controls --- a local machine, not a cloud endpoint --- or it does not run at all.

Annotation-overlap retrieval narrows the differential but leaves three decision-support gaps that matter at the bedside. First, retrieval models score candidates against HPO sets as unordered bags; they do not reason about \emph{causal structure} among a patient's observed phenotypes, even though clinical reasoning routinely uses such structure (``the hyperphenylalaninemia and the intellectual disability are downstream of the same enzymatic defect''). Second, once a shortlist is produced, the clinician receives no signal for \emph{which candidate to trust}: a well-scored top-1 can be wrong for reasons invisible to the scorer --- e.g., a 100\%-obligate pathognomonic feature is absent --- and the clinician must audit manually. Third, retrieval models are single-shot; they cannot tell the clinician \emph{what to measure next} to reduce the differential, even though in practice one additional biomarker or exam finding routinely collapses the shortlist. These are not limitations of scale --- they are limitations of framing. We argue that closing them requires treating rare disease diagnosis as a \emph{multi-agent, closed-loop, causally-grounded retrieval-plus-reasoning problem}, not a single scoring function.

CausalOdyssey is that reframing. A classical IC-weighted gene-rescue backbone produces a top-30 slate; four local agents, each calling Gemma~4~26B, then process the slate in sequence. A \emph{Diagnostician} reranks the top-30 to a top-5 using chain-of-thought over patient-specific FCI causal edges~\cite{chen_igfci_plos_compbio_2022,spirtes_glymour_scheines_2000} and per-candidate Reactome pathway context~\cite{reactome}. A \emph{Verifier} deterministically audits the top-1 against pathognomonic HPO annotations and a MONDO-bridged~\cite{mondo} chain-closure check, and Gemma turns that evidence into a \{confirm, flag, refute\} verdict with a 4.2$\times$ confirm-to-refute discrimination in top-1 accuracy. An \emph{Interrogator} selects the single next HPO whose presence or absence would most change the shortlist, via a deterministic split-variance score weighted by information content (Gemma writes only its rationale). A \emph{Narrator} emits citation-validated clinical prose in which every claim resolves to a trace key. All four agents run on one H100 via Ollama; no patient data leaves the machine. Only the Diagnostician moves candidates; the other three agents add a verdict, a follow-up proposal, and prose to its ranking. On RareBench $n{=}1{,}122$, reranking a top-30 slate lifts pooled $R@1$ by $+12.57$~pp against our structural baseline ($p{=}8.3{\times}10^{-24}$), though a simpler no-LLM rescoring of the same candidates reaches a similar pooled $R@1$ with a very different per-corpus profile (\S\ref{sec:results}), while exposing a structural ceiling (30\% of patients have gold past rank 30 and cannot be rescued by any reranker) that prior work rarely reports.

\textbf{Contributions.}
\begin{itemize}
    \item A local four-agent pipeline on an open-weight model whose mechanism-grounded Diagnostician rerank lifts pooled $R@1$ by +12.57\,pp over our structural baseline on $n{=}1{,}122$, with a comparison against a stronger no-LLM rescoring baseline showing the gain is corpus-dependent.
    \item A per-patient FCI causal-structure layer feeding Gemma's chain-of-thought, inspired by the individualized causal-inference framing of~\cite{chen_igfci_plos_compbio_2022}.
    \item A Verifier that grounds a Gemma verdict in a deterministic chain-closure audit over Reactome + MONDO, giving a 4.2$\times$ rejection-signal discrimination (P(correct\,|\,confirm)=51.4\% vs P(correct\,|\,refute)=12.1\%).
    \item A masked-phenotype protocol for evaluating next-test suggestions that avoids annotation-oracle circularity, with a ceiling (+11\,pp $R@1$ for a perfect follow-up) and a negative result for an annotation-separating heuristic.
    \item An evaluation caution from our own runs: agent-removal reruns moved $R@1$ by up to 2.4\,pp although the removed agents cannot change the ranking, so single-run differences of this size reflect LLM rerun variance, not agent contributions (\S\ref{sec:results:ablation}).
    \item Fully local inference (no cloud APIs) with public code, data, and a live demo.
\end{itemize}

\section{Related work}\label{sec:related}

\textbf{HPO-based retrieval for rare disease.} Phen2Gene~\cite{phen2gene} ranks candidate genes by summing per-term gene scores from a precomputed HPO-to-gene knowledgebase (built from HPO annotations~\cite{hpoa} and OMIM/Orphanet/ClinVar gene--disease links), treating the phenotype set as an unordered bag. LIRICAL~\cite{lirical} computes per-disease likelihood ratios, multiplying per-term ratios under a conditional-independence assumption. PubCaseFinder~\cite{pubcasefinder} adds case-report text mining. Phrank~\cite{phrank} and related IC-weighted similarity measures formalize the bag-of-HPOs intuition with information-content scoring. All four approaches share an architecture: a scoring function maps (patient HPO set, candidate gene or disease) to a real number, and the shortlist is the top-$k$ by score. None reason about dependencies among the patient's observed phenotypes, audit the top-1 against pathognomonic annotations after the fact, or suggest the next test. Our structural backbone is explicitly an IC-weighted gene-rescue retriever; the contribution sits in what the agents add on top of it, not in replacing the retriever.

\textbf{LLMs for rare disease diagnosis.} RareBench~\cite{rarebench} assembles $n{=}1{,}122$ public cases across four corpora (RAMEDIS, LIRICAL, HMS, MME) and evaluates GPT-4, Gemini Pro, GLM-4 and open 7B models on HPO-based diagnosis. Recent work~\cite{gpt4_rare_disease} evaluates GPT-4 as a differential-diagnosis reasoner on case vignettes, finding strong sensitivity to prompt format and model version. PhenoGPT~\cite{phenogpt} addresses the upstream step, fine-tuning language models to extract phenotype terms from clinical free text. The diagnostic efforts measure single-shot accuracy on a ranking task, comparing the LLM's top-$k$ to the gold label; none evaluate next-test suggestions, a discriminative rejection signal, or an explicit decomposition between the retrieval-coverage ceiling and the rerank-within-retrievable lift. RareBench in particular notes that open models trail GPT-4 by a wide margin at $R@1$; our local Gemma~4~26B pipeline narrows that gap on RAMEDIS without any cloud API.

\textbf{Individualized causal discovery.} The FCI algorithm~\cite{spirtes_glymour_scheines_2000} and its modern implementations~\cite{causallearn} recover Partial Ancestral Graphs in the presence of unobserved confounders at the population level. The standard use is population inference: pool a cohort's measurements, learn a single causal graph. Chen et al.~\cite{chen_igfci_plos_compbio_2022} propose an individualized causal-inference framework (iGFCI) that learns one graph per instance by conditioning the search's independence tests on the target instance's own variable values while using the cohort as training data, originally applied to tumor-microenvironment intercellular communication networks. We borrow the individualized framing but not the algorithm: we run standard FCI on a per-patient matrix, and instead of using the recovered graph as a final product for clinical interpretation, we treat each per-patient run as a source of structured context for an LLM reranker. Bidirected edges (shared latent mechanism) become a prompt-time signal that HPO pairs are co-caused, which guides chain-of-thought reasoning over the candidate differential. To our knowledge, per-patient causal discovery has not previously been used as LLM context in medical retrieval.

\textbf{Multi-agent LLM systems and calibrated rejection.} Recent work on agent composition --- ReAct-style scratchpads, generator-critic pairs, and verifier-augmented reasoning --- demonstrates that layering specialized agents can lift accuracy beyond what a single LLM call achieves. In medical applications, however, multi-agent work has mostly targeted single-turn accuracy on closed-book QA. Two design choices distinguish our pipeline. First, the Verifier's evidence is \emph{deterministic}: it audits the top-1 against the pathognomonic annotation set and a chain-closure graph over Reactome/MONDO, and the LLM that issues the verdict sees that evidence rather than re-deriving it. The verdict produces the 4.2$\times$ confirm-to-refute discrimination we report in \S\ref{sec:results}. Second, we evaluate the next-question step directly. The closest prior work in medical LLMs focuses on multi-turn dialogue~\cite{amie_tu_2025} and reports aggregate diagnostic accuracy against conversation length, not the value of a single chosen next question. We propose a masked-phenotype protocol for that question and report that our heuristic fails it (\S\ref{sec:closedloop}).

\section{Method}\label{sec:method}

\begin{figure}[t]
\centering
\includegraphics[width=0.88\linewidth]{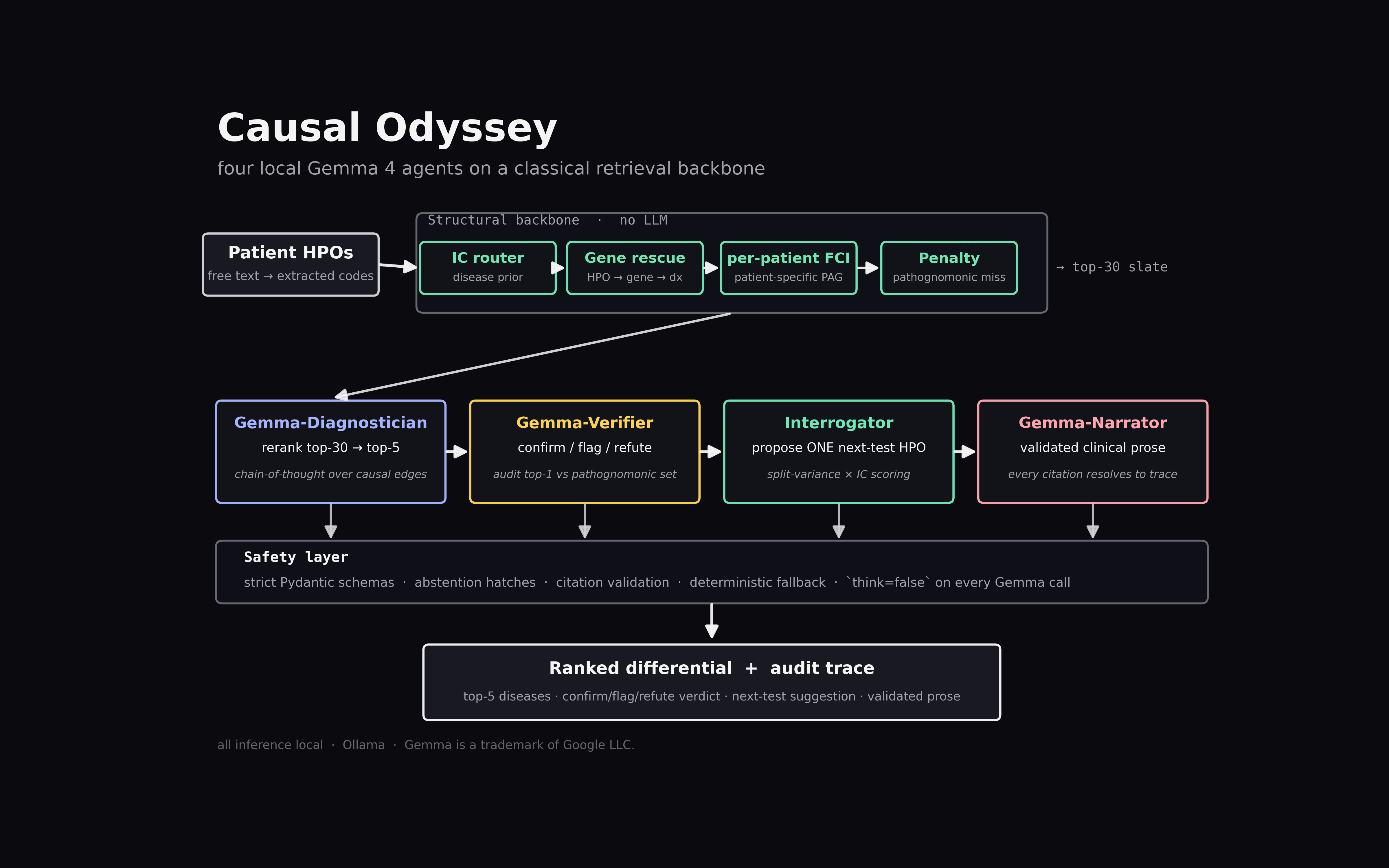}
\vspace{-0.5em}
\caption{\textbf{CausalOdyssey architecture.} A no-LLM backbone (IC router $\to$ gene rescue $\to$ per-patient FCI~\cite{spirtes_glymour_scheines_2000}~$\to$ pathognomonic-miss penalty) returns a top-30 slate. Four agents run in sequence, each calling Gemma~4~26B (the Verifier's evidence and the Interrogator's HPO selection are deterministic; only the Diagnostician changes the ranking), with strict Pydantic schemas, abstention hatches, citation validation, and deterministic fallback. All inference is local via Ollama.}
\label{fig:arch}
\end{figure}

\subsection{Structural backbone}\label{sec:method:backbone}
The structural backbone is a deterministic retrieval pipeline that runs in under 500~ms per patient and produces the top-30 disease slate that the Gemma agents consume (Figure~\ref{fig:arch}). It has four components in series: IC routing, gene rescue, per-patient FCI, and a pathognomonic-miss penalty.

\textbf{IC routing.} A patient-level router decides whether to filter low-information terms. If the patient's HPO set contains a mortality term (e.g.\ \texttt{HP:0001522} Death in infancy; in RareBench these are disease-annotation artifacts rather than observations) or fewer than $15\%$ of its terms have $\mathrm{IC} \ge 6$, all terms with $\mathrm{IC} < 5$ are dropped; otherwise the set is used unchanged. The filter fired for 780 of 1{,}122 patients (559 of 624 in RAMEDIS). Filtered terms are removed from every downstream step (causal discovery, retrieval, and all agent prompts). The router is deterministic.

\textbf{Gene rescue.} Each patient HPO $h \in P$ votes, through the HPOA phenotype-to-gene table, for every (gene, disease) pair it is linked to, weighted by $\mathrm{IC}(h)$ and down-weighted by how many diseases the gene causes; votes through genes shared by both ends of a patient FCI bidirected edge (\S\ref{sec:method:fci}) are multiplied by $1.5$. The disease score is normalized by the number of genes $G_d(P)$ through which $P$ reached $d$:
\[
\mathrm{Score}(d) \,=\, \frac{1}{|G_d(P)|}\sum_{h \in P}\ \sum_{g:\,(h,g,d)\in\mathrm{HPOA}} \frac{\mathrm{IC}(h)\,\beta_g}{|\mathrm{diseases}(g)|},
\]
where $\beta_g = 1.5$ if $g$ links both ends of a bidirected edge and $1$ otherwise, and $P$ is the routed HPO set. The inverse disease-count weight and the gene-count normalization keep genes and disease families with many associations from dominating retrieval. The structural baseline therefore already uses the per-patient causal graph, through $\beta_g$.

\textbf{Missing-pathognomonic penalty.} A candidate disease $d$ is penalized when the patient's HPO set lacks a pathognomonic feature of $d$, defined as an HPO annotated to $d$ at Orphanet frequency $\ge$ \emph{Very frequent} (80--99\%) or \emph{Obligate} (100\%) and with $\mathrm{IC} \ge 8.0$. For each candidate in the top-30, let $\mathrm{pen}(d) = \sum_{h \in \mathrm{missing}(d)} \mathrm{IC}(h)\, w_d(h)$, where $w_d(h)$ is the frequency weight ($3$ for \emph{Very frequent}, $5$ for \emph{Obligate}). The score is multiplied by $\exp\!\big(-\alpha\, \mathrm{pen}(d) / \max_{d'} \mathrm{pen}(d')\big)$ with $\alpha = 0.5$, the maximum taken over the slate, so the penalty reorders candidates within the top-30. This pushes down diseases for which the patient's chart lacks disease-defining findings --- the retrieval analogue of a clinician saying ``this can't be X because the patient doesn't have Y, which is obligate for X.''

\textbf{Top-30 cutoff.} We retain a top-30 slate for reranking. Two distinct cutoffs must not be confused. The first is the gene-rescue \emph{pool depth}: at pool depth 30 the structural backbone achieves $69.70\%$ $R@30$ on RareBench $n{=}1{,}122$ --- this is the retrieval-coverage ceiling (\S\ref{sec:limit}), and pushing it requires upstream gene-rescue work, not deeper pools, because the missing patients are largely absent from the reachable set. The second is the \emph{Diagnostician slate width}: how many of the pooled candidates the LLM actually reranks. This width is a recall--precision dial that we sweep in \S\ref{sec:results:slate}; $R@1$ peaks at width 30. We fix both to 30 for the headline results.

\subsection{Per-patient causal structure}\label{sec:method:fci}
For each patient we construct an individualized disease-by-HPO binary matrix and run Fast Causal Inference (FCI) to recover a Partial Ancestral Graph (PAG) over the patient's observed phenotypes. This follows the individualized framing of Chen et al.~\cite{chen_igfci_plos_compbio_2022} in spirit --- one graph per instance --- but uses standard FCI~\cite{spirtes_glymour_scheines_2000} as implemented in causal-learn~\cite{causallearn} on the per-patient matrix, without iGFCI's instance-conditioned independence tests.

\textbf{Matrix construction.} Rows are diseases that share at least one HPO with the patient (median 158 per patient, interquartile range 72--985). Columns are the patient's routed HPOs that vary across those diseases (median 5, IQR 3--9). Each cell $(d, h)$ is 1 if $h$ is annotated to $d$ in the expanded disease database, else 0. FCI is skipped when fewer than 20 diseases match (216 patients) and yields no graph when fewer than two columns vary (50 patients).

\textbf{FCI parameters.} We run FCI with the chi-square conditional-independence test at $\alpha = 0.05$, no prior background knowledge, and show-progress disabled. A wall-clock cap of 60~s is applied per patient; on timeout the pipeline falls back to the score-based GES variant from causal-learn. In the slate-30 run the timeout never fired.

\textbf{PAG edge semantics.} The resulting graph has five edge types, encoded in causal-learn as the adjacency-matrix pair $(\text{graph}[i,j], \text{graph}[j,i])$, where $\text{graph}[i,j]$ is the endpoint mark at node $i$ ($-1$ tail, $1$ arrowhead, $2$ circle):
\[
\underbrace{(-1, 1)}_{i\to j}\ \ \ \underbrace{(1, 1)}_{i\leftrightarrow j}\ \ \ \underbrace{(2, 1)}_{i\,\circ{\to}\,j}\ \ \ \underbrace{(2, 2)}_{i\,\circ{-}\circ\,j}\ \ \ \underbrace{(-1, -1)}_{i - j}.
\]
The key edge type for our purposes is the \emph{bidirected} edge $i \leftrightarrow j$: it indicates that the pair of HPOs share a common unobserved cause, which in this domain typically means a shared upstream genetic or enzymatic lesion. Directed edges $i \to j$ carry weaker information (phenotype-level causation can reflect anatomical dependency rather than mechanism).

\textbf{What Gemma consumes.} Only the bidirected edges (up to ten) are passed to the Diagnostician, as ``shared-mechanism phenotype pairs''; directed, partially oriented, and circle edges are not shown. In the slate-30 run, 264 of 1{,}122 patients ($23.5\%$) had at least one bidirected edge; for the rest the block reads ``none''. Whether these edges help the rerank has not been tested.

\subsection{Mechanism graph}\label{sec:method:mechgraph}
The Mechanism Graph is a pre-computed, static knowledge graph over genes, Reactome pathways, and MONDO disease concepts that provides a deterministic \emph{chain-closure audit} independent of the LLM. It is the structural signal the Verifier uses to check whether a Gemma-cited pathway actually supports a Gemma-predicted disease.

\textbf{Construction.} We build a per-gene cache by querying Reactome's REST API with HGNC-normalized gene symbols, mapped through UniProt where necessary. Each entry stores the pathways the gene participates in, each pathway's \texttt{isInDisease} flag, and the pathway's disease cross-references (DOID, OMIM, Orphanet, MONDO), which are mapped to MONDO when the audit runs. The cache holds one JSON record per gene and covers 3{,}967 of the 5{,}250 HPOA-annotated genes ($75.6\%$); the uncovered 1{,}283 are primarily non-coding RNAs and understudied genes for which Reactome has no pathway entry. MONDO disease concepts are cross-referenced to OMIM, Orphanet, and ICD-10 through the MONDO SSSOM mappings.

\textbf{Four closure categories.} For a candidate disease $d$ with supporting genes $G_d$ and the Reactome pathways $r$ Gemma cites for it, the audit assigns one of four levels (the strongest over the cited pathways):
\begin{itemize}
    \item \emph{Tight-Mondo}: $r$ shares a MONDO identifier with $d$ (via the pathway's disease cross-references). This is the strongest machine-checkable signal --- the pathway is explicitly disease-specific.
    \item \emph{Tight-isInDisease}: $r$ is flagged \texttt{isInDisease} by Reactome and contains at least one gene in $G_d$, but no direct MONDO cross-reference to $d$. Disease-specific by annotation, just not resolved through MONDO.
    \item \emph{Loose}: $r$ contains a gene in $G_d$ but is not disease-specific (e.g., a general metabolism pathway that happens to include the supporting gene). Circumstantial.
    \item \emph{None}: no cited pathway belongs to a supporting gene of $d$ (including when nothing is cited); pathways that are not on a supporting gene are ignored. The Verifier passes the level to Gemma; no category triggers a deterministic refutation.
\end{itemize}

\textbf{Coverage and calibration.} In the slate-30 RareBench run ($n{=}1{,}119$ patients with an audited top-1) we observe $35.9\%$ tight closure (402 patients), $39.2\%$ loose, and $24.8\%$ with no closure. The four-category resolution is deliberately coarse; finer-grained semantic scoring (e.g., pathway-to-phenotype specificity) is left for future work. The audit is model-agnostic: it operates purely on Gemma's emitted pathway citations and the static graph, so it can be re-run offline on any Gemma output without re-invoking the model.

\subsection{Gemma agent stack}\label{sec:method:agents}

\paragraph{Diagnostician.}
The Diagnostician takes the top-30 structural slate and reranks it, with per-candidate confidence, a short reason, and pathway and PubMed citations. It is the only agent whose output determines the reported ranking and hence $R@k$. The prompt (\texttt{MECHANISM\_RERANK\_PROMPT}, Appendix~\ref{app:prompts}) contains two retrieved worked examples (\S\ref{sec:eval}), the patient's routed HPOs with names, the bidirected FCI pairs, and, for each of the 30 candidates, its name, supporting genes, a sample of its annotated phenotypes, and a Reactome block for up to three supporting genes listing pathway IDs, names, and whether each is disease-specific. Gemma returns a 0--100 score per candidate with a short reason and optional Reactome and PubMed citations, which must come from that candidate's block. The final ranking is deterministic given those scores:
\[
\mathrm{final}(d) = 0.6\,\frac{\mathrm{Score}(d)}{\max_{d'}\mathrm{Score}(d')} + 0.4\,\frac{\mathrm{gemma}(d)}{100} + b(d),
\]
where $b(d) = 0.15$ if a cited pathway gives tight chain closure (\S\ref{sec:method:mechgraph}) and $0.05$ if loose. The whole slate is reranked, so $R@10$ and $R@30$ are defined; the candidate's confidence is its Gemma score scaled to $[0,1]$, and the remaining \texttt{DiagnosticianOutput} fields (Appendix~\ref{app:schemas}) are filled by code. Inference: temperature 0.3, $N{=}1$, \texttt{num\_predict} 3600 at slate width 30, JSON mode, \texttt{think=false} (required for Gemma~4 reasoning-token handling).

\paragraph{Verifier.}
The Verifier has two stages. A deterministic stage computes evidence against the Diagnostician's top-1 from static annotation data; a Gemma call then reads that evidence and issues the verdict. The evidence is never produced by the model that made the prediction, but the verdict is: the Verifier's Gemma call uses the same model as the Diagnostician, in a separate call with a separate prompt (Appendix~\ref{app:prompts:verifier}). The deterministic stage runs three checks.

\emph{Check 1 (pathognomonic presence).} Enumerate HPOs annotated to top-1 with $\mathrm{IC} \ge 8$ at frequency $\ge$ \emph{Very frequent} (80--99\%) in HPOA. Any that are \emph{not} in the patient's HPO set are flagged as \texttt{expected\_but\_missing}. \emph{Check 2 (unexpected-present).} Conversely, patient HPOs that are \emph{not} in top-1's annotation set are collected as \texttt{unexpected\_present}. \emph{Check 3 (chain-closure audit).} Using \S\ref{sec:method:mechgraph}'s graph, classify top-1's cited pathways as tight-MONDO, tight-\texttt{isInDisease}, loose, or none.

Gemma receives the top-1, its supporting genes, the patient's HPOs, and the three check outputs, and returns \texttt{confirm} (biology consistent), \texttt{flag} (partial fit), or \texttt{refute} (strong mismatch, e.g.\ an obligate phenotype definitively absent), with a short rationale. The rationale may cite only HPOs present in the evidence; if it fails that check, or the call fails, no verdict is issued. The verdict is not a fixed function of the check outputs: patients with the same counts of missing and unexpected phenotypes receive all three verdicts. It enters the confidence fusion as a multiplier (v2, below) and does not change the ranking.

\emph{Calibration.} On the $n{=}1{,}108$ patients in our slate-30 RareBench eval with a Verifier verdict, the trichotomy distributes as \{confirm: $57.6\%$, flag: $29.8\%$, refute: $12.6\%$\} (\S\ref{sec:results}). The conditional probabilities $\mathrm{P}(\text{correct}\,|\,\text{confirm}) = 51.4\%$ vs $\mathrm{P}(\text{correct}\,|\,\text{refute}) = 12.1\%$ reflect a $4.2\times$ confirm-to-refute discrimination --- the rejection signal is discriminative enough that a \emph{refute} verdict should substantially shift a clinician's prior.

\emph{Confidence-fusion v2.} An earlier iteration (v1) fused the Diagnostician confidence, an additive verdict adjustment ($\Delta c \in \{0.0, -0.1, -0.3\}$ for confirm/flag/refute), the Narrator validation rate, and a structural-anchor term with equal weights; AUC on the ``top-1-correct'' binary task was 0.605 (the structural anchor dragged signal --- AUC 0.514 alone --- and the Narrator validation rate added noise at AUC 0.636; v1 figures measured on the slate-10 run). The v2 fusion uses $c = c_{\text{diag}} \cdot m_{\text{verdict}}$ where $m \in \{1.0, 0.5, 0.3\}$ for confirm/flag/refute (capped at $0.70$ if an agent abstains; when Gemma issues no verdict, $m$ is inferred from the deterministic flags); at the slate-30 headline it achieves AUC $0.780$ (vs.\ $0.726$ for Diagnostician confidence alone). The v2 simplification is the configuration we ship; v1 is retained in the codebase for backward compatibility but not used in the headline results.

\paragraph{Interrogator.}
One next-test HPO is selected deterministically (Alg.~1) over the top-5 candidates (the Diagnostician's in the pipeline; the structural top-5 in the offline evaluation of \S\ref{sec:closedloop}); a separate Gemma call then writes a short rationale for the chosen HPO and does not change the choice.

\begin{figure}[h]
\small
\noindent\hrulefill
\textbf{Algorithm 1.} Interrogator --- next-test HPO selection.
\vspace{-0.5em}
\begin{flushleft}
\textbf{Input:} patient HPOs $P$; top-5 candidates $C_1..C_5$; annotation frequencies $p_i(h)$ (Obligate $1.0$, Very frequent $0.85$, Frequent $0.55$, Occasional $0.17$, Very rare $0.02$, unannotated $0$); FCI bidirected edges $E_b$.\\
\textbf{Output:} proposed HPO $h^{*}$, or abstain.\\[0.3em]
\textbf{1.}\ \textbf{for each} $h \in \big(\bigcup_i \mathrm{ann}(C_i)\big) \setminus P$ with $\mathrm{IC}(h) > 0$ \textbf{do}\\
\hspace*{1.5em} \textbf{skip} unless some $p_i(h) \ge 0.3$ and some $p_i(h) \le 0.2$\\
\hspace*{1.5em} $s \gets \mathrm{sd}_i\big(p_i(h)\big) \cdot \mathrm{IC}(h)$\\
\hspace*{1.5em} \textbf{if} $h$ is bidirected-linked to some $h' \in P$ via $E_b$ \textbf{then} $s \gets 0.6\,s$ \ \ \textit{(redundancy penalty)}\\
\hspace*{1.5em} \textbf{if} $h \in \mathrm{ann}(C_1)$ \textbf{then} $s \gets 1.3\,s$ \ \ \textit{(top-1 boost)}\\
\textbf{2.}\ $h^{*} \gets \arg\max_h s$; \textbf{abstain if} $s(h^{*}) < 0.25$
\end{flushleft}
\vspace{-0.3em}
\noindent\hrulefill
\end{figure}

\emph{Intuition.} The spread of $p_i(h)$ across the five candidates measures how well observing $h$ would split the shortlist: a phenotype expected in some candidates and absent in others separates them, one common to all does not. Multiplying by IC favors specific findings over common ones. The $0.6$ redundancy penalty and $1.3$ top-1 boost were set by hand and were not tuned on the evaluation set (or on any held-out set); we report no sensitivity analysis for them.

\emph{Abstention.} The Interrogator abstains when the best score is below $0.25$ or no phenotype passes the filter. In the slate-30 run this happened for 1 of 1{,}122 patients.

\paragraph{Narrator.}
The Narrator produces a short clinician-readable summary over the accumulated \texttt{DiagnosisTrace}. It is the only agent whose output is prose rather than structured fields, and it is subject to a \emph{citation-validation} step: every HPO, gene, or disease identifier mentioned in the generated narrative must resolve to an entry in the trace. If validation fails (\texttt{src/agents/validators.py}), the Narrator is retried once with the validator's feedback; if it still fails, the whole narrative is replaced by a deterministic template composed from trace fields.

\emph{Validation rates.} On RareBench $n{=}1{,}122$ (slate-30 run), $24.9\%$ of patients (279) receive a Gemma-generated narrative that passes the citation check; the remaining $75.1\%$ fall back to the deterministic template. The low pass rate reflects Gemma's tendency to synthesize plausible-sounding but non-trace-grounded citations (e.g., inventing a PMID or extrapolating from a related disease); the validator catches these. We report the pass rate in the results section as an interpretability measure rather than a performance metric.

\emph{The Narrator does not affect ranking.} The Narrator runs after the ranking is fixed and its confidence is not part of the shipped fusion (v2, defined under Verifier above), so it cannot change $R@k$; it is a clinician-facing output only.

\subsection{Safety layer}\label{sec:method:safety}
Four independent guards wrap every Gemma call. First, model outputs are parsed leniently (JSON salvage) and code then builds Pydantic objects (Appendix~\ref{app:schemas}); if the Diagnostician call fails, a structural stub ranking is used. Second, every agent exposes an \texttt{abstained: bool} field and an \texttt{abstain\_reason: str}; downstream agents read and respect abstentions rather than pretending a valid output was produced. Third, Narrator claims must resolve to trace keys (the trace is accumulated by the upstream stages, \S\ref{sec:method:backbone}--\S\ref{sec:method:agents}, and validation checks against it). Fourth, all Gemma calls set \texttt{think=false} in the Ollama request body. This is a Gemma~4~26B-specific requirement: the model has built-in thinking capability, and without explicit \texttt{think=false} the reasoning tokens are routed to the \texttt{.thinking} field and \texttt{.response} returns empty. We encountered this as a silent failure early in development and now apply it universally to every Gemma integration. Together, the four guards ensure that when Gemma fails --- malformed JSON, truncated output, hallucinated citation, or empty response from missing \texttt{think=false} --- the pipeline degrades gracefully to a deterministic fallback rather than crashing or producing silently-wrong output.

\section{Evaluation setup}\label{sec:eval}

\textbf{Dataset.} We evaluate on the RareBench unified benchmark~\cite{rarebench}, $n{=}1{,}122$ patients across four corpora drawn from published rare-disease case repositories. RAMEDIS ($n{=}624$) draws from the RAMEDIS rare metabolic diseases database and biases toward pediatric inborn errors of metabolism. LIRICAL ($n{=}370$) is RareBench's filtered subset of the 384 published case reports (262 Mendelian diseases) used to validate LIRICAL~\cite{lirical}, and spans a broader diagnostic range than RAMEDIS. HMS ($n{=}88$) is drawn from the Hannover Medical School outpatient clinic for rare inflammatory systemic diseases~\cite{ronicke_adadx_2019} (93 cases in the original study, 88 after curation into HPO/OMIM--Orphanet form by the PhenoBrain group~\cite{phenobrain_mao_2025}), so it skews toward autoinflammatory, vasculitic, and rheumatologic cases rather than monogenic disease. MME ($n{=}40$) is from Matchmaker Exchange and consists of ultra-rare cases with small cohort sizes per disease. (RareBench additionally includes an access-restricted clinical corpus, PUMCH; we evaluate on the four public configurations, which total exactly $n{=}1{,}122$.) Each patient has an HPO-code set and a gold disease ID.
Cohort sizes reflect the underlying corpora; we do not rebalance across datasets because doing so would obscure the per-domain performance that our analysis hinges on.

\textbf{Gold-match logic.} RareBench gold labels are heterogeneous: a single patient may be annotated with OMIM, Orphanet, or CCRD identifiers that cross-reference the same disease. We accept a prediction if it equals a gold identifier, maps to it through the Orphanet--OMIM crosswalk, or falls in the same MONDO family, defined as sharing a direct MONDO parent (one parent per term). Family matching is more lenient than strict matching: under strict matching (exact identifier or crosswalk) FullMA's $R@1$ is $25.6\%$ against $13.9\%$ for the structural baseline ($+11.7$~pp), compared with $35.3\%$ and $22.7\%$ family-aware. We report family-aware throughout. The resolver is used only for evaluation; the Verifier never sees the gold label.

\textbf{Metrics.} We report $R@1$, $R@5$, $R@10$ --- the fraction of patients whose gold disease appears in the reranker's top-$k$. Pairwise significance is assessed by McNemar's exact test on same-patient top-1 decisions; the paired design is appropriate because we are comparing arms on identical patients rather than sampling. Confidence intervals on $\Delta R@K$ are 95\% percentile intervals from a paired patient-level bootstrap (10{,}000 resamples with replacement, NumPy \texttt{default\_rng(42)}; each resample draws the same patients for both arms). AUC confidence intervals use 2{,}000 resamples, and the masked-phenotype intervals (\S\ref{sec:closedloop}) resample patients rather than patient-seeds (2{,}000 resamples). We use McNemar rather than a $t$-test because the outcome is binary per patient and the $t$-test overstates significance on small effect sizes with binary outcomes.

\textbf{Arms.} \emph{Baseline} is the structural backbone only (\S\ref{sec:method:backbone}), with no LLM involvement; this isolates the retrieval-coverage ceiling from any reranking lift. \emph{FullMA} is the full pipeline of \S\ref{sec:method:agents}; we report it at slate widths 10 and 30 (headline). \emph{A3} is an earlier single-agent run (Diagnostician + per-patient FCI + chain-of-thought + mechanism context, no downstream agents) that exists only at slate width 10. Because only the Diagnostician determines the ranking, FullMA and A3 differ in $R@k$ only through their Diagnostician calls: A3 used an earlier revision of the Diagnostician prompt (its outputs carry no PubMed citations; FullMA's do), and both are single samples at temperature 0.3. The A3-vs-FullMA comparison is therefore a comparison of Diagnostician prompt revisions, not a measure of what the downstream agents add.

\textbf{Hardware and reproducibility.} Gemma~4~26B (quantized) runs via local Ollama on a single NVIDIA H100. All Gemma calls use $N{=}1$ sample and \texttt{think=false}. The Diagnostician uses temperature 0.3 and \texttt{num\_predict} 3600 at slate width 30; the Verifier temperature 0.2 (300 tokens); the Interrogator rationale temperature 0.3 (300); the Narrator temperature 0.2 (1200, one retry).

\textbf{Few-shot exemplars come from the evaluation cohort.} The Diagnostician prompt includes two worked examples, retrieved per patient as the most similar entries (by overlap of structural candidate sets) in a bank of 200 RareBench patients (140 RAMEDIS, 20 each from LIRICAL, HMS, MME), each shown with its correct diagnosis. A leave-one-out guard removes the patient itself and any bank entry whose diagnosis ID equals one of the patient's gold IDs, but the bank is drawn from the evaluation set and the guard does not remove MONDO-family equivalents. Re-deriving the exemplars for the slate-30 run, $51$ of $1{,}122$ patients ($4.5\%$) were shown an example whose answer counts as their gold under our family-aware matching. These patients did not benefit (FullMA $R@1$ $15.7\%$ vs $36.2\%$ for the rest), and excluding them leaves the headline lift unchanged ($+12.98$~pp, 170 gain vs 31 lose, $p{=}2.0{\times}10^{-24}$, $n{=}1{,}071$). The results are nonetheless not fully held out; a run without the bank and an external cohort would be needed to remove this. Structural backbone inference is fully deterministic; Gemma inference is not, and reruns may produce slightly different top-5 orderings on the boundary cases. All data (RareBench, HPOA, MONDO, Reactome, Orphanet), the demo code (MIT license), prompts, and cached Gemma outputs for the showcase cases are public; the evaluation code differs from the public demo version (\S\ref{sec:availability}).

\section{Results}\label{sec:results}

\begin{figure}[h]
\centering
\includegraphics[width=0.82\linewidth]{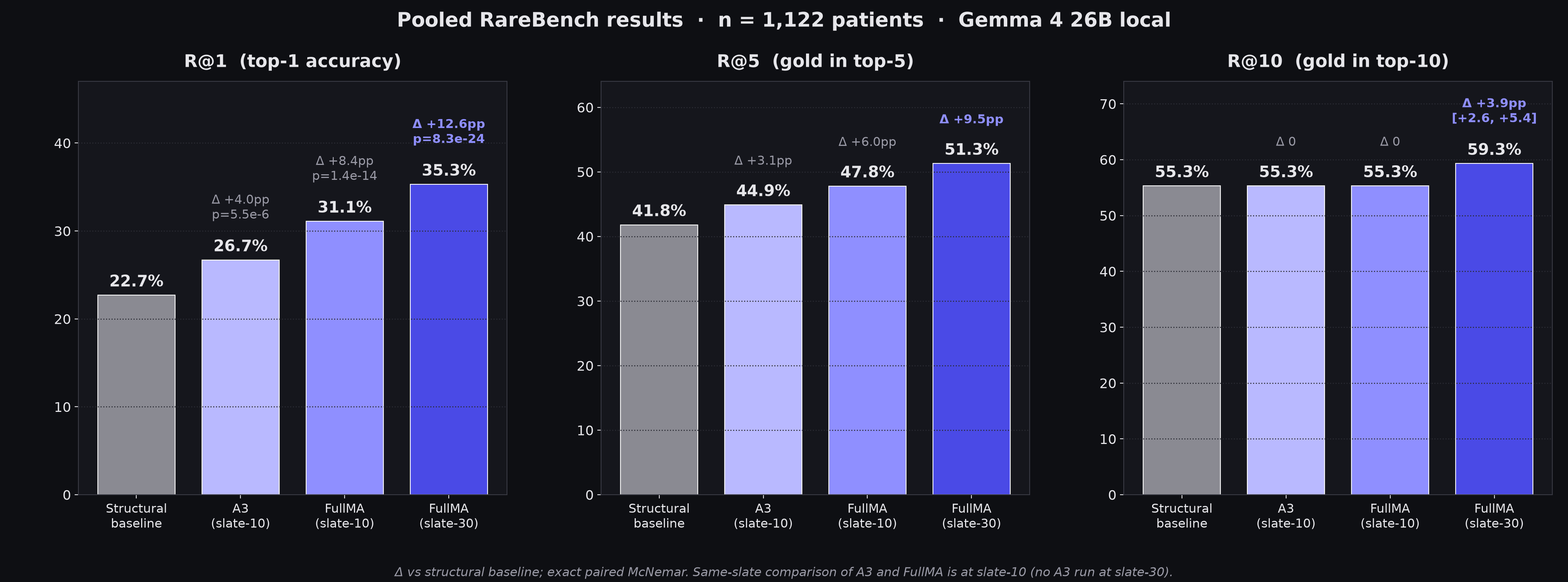}
\vspace{-0.5em}
\caption{\textbf{Pooled R@K on RareBench} ($n{=}1{,}122$, paired). At slate-30, FullMA lifts $R@1$ from 22.7\% to 35.3\% (+12.57\,pp, exact McNemar $p{=}8.3{\times}10^{-24}$) and $R@10$ by +3.92\,pp. A3 exists only at slate-10; its same-slate comparator is FullMA at slate-10. $R@30$ (69.7\%, not shown) is identical across arms: the gene-rescue pool depth sets the retrieval-coverage ceiling.}
\label{fig:headline}
\end{figure}

\subsection{Pooled results}
Table~\ref{tab:pooled} and Figure~\ref{fig:headline} report paired R@K with 95\% bootstrap CIs on RareBench $n{=}1{,}122$.

The structural baseline recovers the gold diagnosis at $R@1 = 22.73\%$, $R@5 = 41.80\%$, $R@10 = 55.35\%$, $R@30 = 69.70\%$. At slate-30, FullMA lifts $R@1$ to $35.29\%$ ($\Delta = +12.57$~pp [+10.07, +14.97], exact McNemar $p{=}8.3{\times}10^{-24}$; 176 patients gain, 35 lose) and $R@5$ to $51.34\%$ ($\Delta = +9.54$~pp [+7.49, +11.68]). At slate-10, FullMA reaches $R@1 = 31.11\%$ ($+8.38$~pp [+6.24, +10.52], $p{=}1.4{\times}10^{-14}$) and A3 reaches $26.74\%$ ($+4.01$~pp, $p{=}5.5{\times}10^{-6}$); the same-slate gap of $+4.37$~pp (89 gain, 40 lose, $p{=}1.9{\times}10^{-5}$) reflects the Diagnostician prompt revision between the two runs (\S\ref{sec:eval}), not the downstream agents. We do not have an A3 run at slate-30, so the headline lift should not be compared with the slate-10 A3 number.

\textbf{A stronger no-LLM baseline.} Removing the inverse disease-count weight from gene rescue (\S\ref{sec:method:backbone}), with no LLM, raises structural $R@1$ to $36.27\%$ (Table~\ref{tab:pooled}; $+13.55$~pp over the production baseline, $p{=}1.9{\times}10^{-22}$). Pooled, FullMA and this variant are indistinguishable ($-0.98$~pp [$-4.10$, $+2.23$], 155 vs 166 discordant patients, $p{=}0.58$), but they succeed on different patients. The LLM rerank is much better on RAMEDIS ($43.6\%$ vs $27.6\%$, $+16.0$~pp, $p{=}4.8{\times}10^{-14}$) and much worse on LIRICAL ($27.0\%$ vs $53.0\%$, $p{=}6.4{\times}10^{-21}$) and MME ($27.5\%$ vs $57.5\%$, $p{=}4.9{\times}10^{-4}$); HMS does not differ ($p{=}0.51$). Part of the variant's LIRICAL advantage comes from annotation leakage, but it persists on uncited cases (\S\ref{sec:limit}). We did not run the Diagnostician on the variant's candidates, so whether combining the two would help is unknown. The headline lift is therefore specific to our production baseline; the value of LLM reranking here depends on the corpus.

\textbf{Established tools.} We ran LIRICAL~\cite{lirical} (phenotype-only) and Phen2Gene~\cite{phen2gene} on the same 1{,}122 patients and scored them with the same matcher (Table~\ref{tab:pooled}). FullMA has higher $R@1$ than LIRICAL ($35.3\%$ vs $28.1\%$; 248 vs 167 discordant patients, $p{=}8.2{\times}10^{-5}$) and than LIRICAL with Orphanet diseases ($30.8\%$, $p{=}0.015$), but lower $R@10$ ($59.3\%$ vs $63.8\%$, $p{=}0.011$) and $R@30$ ($69.7\%$ vs $78.2\%$). The per-corpus pattern matches the no-IDF comparison: FullMA is far ahead on RAMEDIS ($43.6\%$ vs $16.7\%$ at $R@1$, $p{=}4.4{\times}10^{-27}$) and behind on LIRICAL ($27.0\%$ vs $50.3\%$) and MME ($27.5\%$ vs $50.0\%$). The LIRICAL corpus is LIRICAL's own validation set and is partly reflected in HPOA (\S\ref{sec:limit}), so the tool's advantage there is not a clean comparison. Phen2Gene ranks genes, not diseases, and is not like-for-like. For context, RareBench reports $32.3\%$ top-1 for GPT-4 on a different case mix that includes the restricted PUMCH corpus~\cite{rarebench}.

\textbf{$R@30$ is the ceiling, not $R@10$.} The flat metric is $R@30 = 69.70\%$: it is identical across arms because it equals the gene-rescue pool depth, and reranking cannot add candidates the pool never retrieved. But $R@10$ is \emph{not} flat at slate-30 --- FullMA lifts it by +3.92~pp [+2.58, +5.35] ($p{=}2.0{\times}10^{-8}$). With a 30-candidate slate the Diagnostician can pull a gold disease sitting at structural rank 11--30 up into the top 10, which a 10-candidate slate structurally cannot. So the rerank does two distinct things: it reorders within the visible window (the $R@1$/$R@5$ lift) \emph{and} it expands the top-10 by promoting deeper candidates (the $R@10$ lift). Only $R@30$ --- the pool depth --- is a hard ceiling, and pushing it requires upstream gene-rescue work, not downstream LLM reasoning.

\textbf{Effect-size context.} A +12.57~pp $R@1$ lift is large by the standards of rare-disease retrieval literature, where the strongest configuration in RareBench~\cite{rarebench} (GPT-4 with knowledge-graph-driven dynamic few-shot prompting) depends on cloud-API inference and no open-weight model in that study approached it. Our lift is achieved with an open-weight model running locally (the 26B-A4B mixture-of-experts variant: 25.2B total parameters, 3.8B active per token), which changes the deployment profile entirely. It is measured against our production structural baseline; a no-LLM variant of that baseline (dropping its inverse disease-count weight) reaches a similar pooled $R@1$ with a different per-corpus profile (\S\ref{sec:limit}).

\begin{table}[h]
\centering\small
\caption{Pooled $R@K$ at $n{=}1{,}122$ (paired). $\Delta$ vs structural baseline with 95\% paired-bootstrap CIs (10{,}000 resamples); McNemar is exact two-sided. A3 exists only at slate width 10 (no slate-30 run). $^\dagger$Structural backbone with the inverse disease-count weight removed, from a structural-only re-implementation that reproduces production ranks for 1{,}048 of 1{,}122 patients; it changes which candidates enter the top-30, so its $R@30$ differs. $^\ddagger$LIRICAL~\cite{lirical} 2.4.1, phenotype-only, HPO/HPOA release 2026-09 (the 2024-04 release gives the same $R@1$); ranks its own disease list. \textsuperscript{\S}Phen2Gene~\cite{phen2gene} ranks genes; rank = best rank of a gene associated with the gold disease. No-LLM rows do not use the slate.\label{tab:pooled}}
\begin{tabular}{lcccccc}
\toprule
Arm & $R@1$ & $R@5$ & $R@10$ & $R@30$ & $\Delta R@1$ [95\% CI] & McNemar $p$ \\
\midrule
Structural baseline & 22.73 & 41.80 & 55.35 & 69.70 & --- & --- \\
A3, slate-10 & 26.74 & 44.92 & 55.35 & 69.70 & $+4.01\,[+2.23,+5.79]$ & $5.5{\times}10^{-6}$ \\
FullMA, slate-10 & 31.11 & 47.77 & 55.35 & 69.70 & $+8.38\,[+6.24,+10.52]$ & $1.4{\times}10^{-14}$ \\
\midrule
No-IDF rescoring (no LLM)$^\dagger$ & 36.27 & 51.96 & 69.34 & 77.54 & $+13.55\,[+10.87,+16.22]$ & $1.9{\times}10^{-22}$ \\
LIRICAL tool, OMIM$^\ddagger$ & 28.07 & 52.50 & 63.81 & 78.16 & $+5.35\,[+2.23,+8.38]$ & $6.6{\times}10^{-4}$ \\
LIRICAL tool, OMIM+Orphanet$^\ddagger$ & 30.75 & 53.03 & 63.19 & 80.12 & $+8.02\,[+4.81,+11.14]$ & $5.1{\times}10^{-7}$ \\
Phen2Gene (gene-level)\textsuperscript{\S} & 22.99 & 37.34 & 43.85 & 54.81 & $+0.27\,[-3.03,+3.57]$ & $0.91$ \\
\midrule
\textbf{FullMA, slate-30} & \textbf{35.29} & \textbf{51.34} & \textbf{59.27} & 69.70 & $\mathbf{+12.57\,[+10.07,+14.97]}$ & $\mathbf{8.3{\times}10^{-24}}$ \\
\bottomrule
\end{tabular}
\end{table}

\subsection{Per-dataset decomposition}
\begin{figure}[h]
\centering
\includegraphics[width=0.82\linewidth]{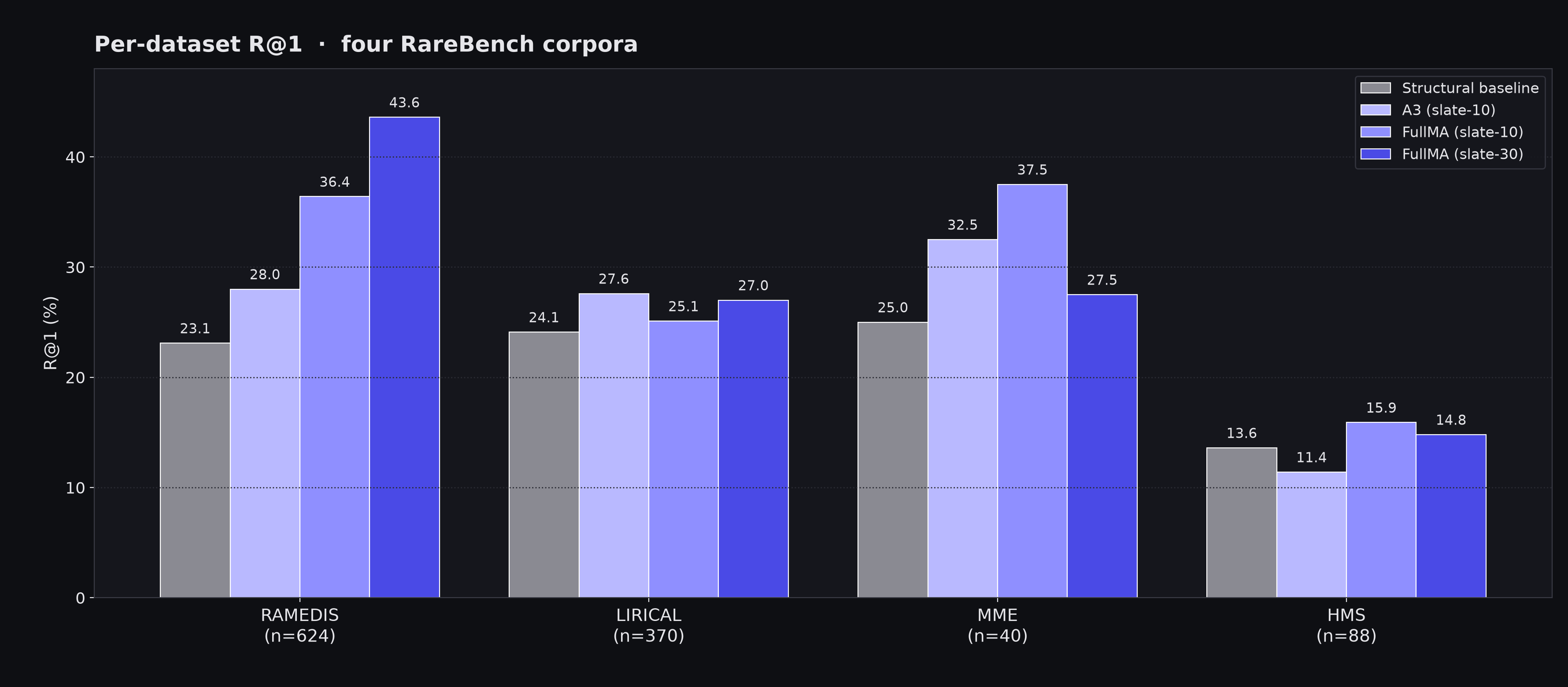}
\vspace{-0.5em}
\caption{\textbf{Per-dataset $R@1$.} FullMA at slate-30 improves on the structural baseline on all four corpora (significant only on RAMEDIS). A3 exists only at slate-10; compare it with FullMA at slate-10.}
\label{fig:perds}
\end{figure}

Figure~\ref{fig:perds} decomposes $R@1$ by corpus. At slate-30, FullMA improves on the baseline in all four datasets, but the lift is significant only on RAMEDIS: RAMEDIS $+20.51$~pp [+16.99, +24.20] ($p{=}2.5{\times}10^{-28}$); LIRICAL $+2.97$~pp (CI [$-0.54$, +6.22], $p{=}0.12$); MME $+2.50$~pp (CI [$-12.50$, +17.50] at $n{=}40$, $p{=}1.0$); HMS $+1.14$~pp (CI [$-4.55$, +6.82] at $n{=}88$, $p{=}1.0$). At the same slate-10 width, FullMA's Diagnostician beats A3's on RAMEDIS ($+8.33$~pp, $p{=}3.1{\times}10^{-9}$), HMS and MME (small $n$, not significant) and trails it on LIRICAL ($-2.43$~pp, 14 gain vs 23 lose, $p{=}0.19$).

\textbf{Why RAMEDIS wins.} RAMEDIS is the in-distribution training domain for our backbone: pediatric metabolic disease with clean one-gene-one-disease Mendelian mapping and dense pathognomonic annotations in HPOA. The gene-rescue retriever produces high-quality top-30 slates (gold in the structural top-10 for $60.4\%$ of RAMEDIS patients and in the top-30 for $72.6\%$); the Diagnostician can then discriminate among metabolically-related differentials using mechanism pathway context (e.g., distinguishing methylmalonic aciduria cblA-type from cblB-type via cited propionylcarnitine pathway). This is the configuration the pipeline was designed for, and the $+20.51$~pp lift reflects that.

\textbf{LIRICAL: small lift, no regression established.} LIRICAL is the corpus where the reranker helps least ($+2.97$~pp at slate-30, $p{=}0.12$), though on the 171 LIRICAL patients whose source paper is not cited by HPOA the lift is $+5.8$~pp ($p{=}0.013$; \S\ref{sec:limit}). At slate-10 the FullMA Diagnostician scores $2.43$~pp below A3's ($p{=}0.19$, not significant). This is not a regression caused by the added agents, since the downstream agents do not change the ranking, and the difference is within the spread we observe between Diagnostician reruns (\S\ref{sec:results:ablation}). LIRICAL's candidate pools contain many phenotypically-similar diseases (dysmorphic syndromes with overlapping HPO profiles), which plausibly makes the single-sample Diagnostician less stable there. Self-consistency ($N{=}3$ samples, majority vote) does not help at the headline slate ($-0.27$~pp, Appendix~\ref{app:sc3}).

\textbf{HMS scope mismatch.} 51\% of HMS patients have gold past top-30 in the structural backbone --- the dataset is dominated by autoimmune and rheumatologic disease with no clean Mendelian gene$\to$disease mapping. Our approach is structurally mismatched to this domain. The $+1.14$~pp lift is a net of one patient (4 gained, 3 lost); the four gains are autoimmune or vasculitic diagnoses (lupus, Beh\c{c}et syndrome, and granulomatosis with polyangiitis twice), so it should not be read as an effect.
We retain HMS in the pooled report for completeness, but the honest framing is that CausalOdyssey's thesis is Mendelian rare-disease retrieval, not autoimmune diagnosis. A separate line of work would need a different backbone for autoimmune cases.

\textbf{MME underpowering.} At $n{=}40$ MME is too small for tight CIs but directionally favorable at $+2.50$~pp (CI [$-12.50$, +17.50]). MME's cohort is ultra-rare cases with very low prior probability; at the wider 30-candidate slate the structural baseline already places gold at rank 1 for $25\%$ of these patients, so the headroom for rerank lift is correspondingly smaller than at the 10-candidate slate.

\subsection{Stratified lift by baseline rank}
Figure~\ref{fig:strat} decomposes FullMA's pooled slate-30 $+12.57$~pp $R@1$ lift by the structural backbone's baseline rank for each patient's gold disease. The decomposition reveals five strata with qualitatively different behavior.

\emph{Rank 1 ($n{=}255$).} Patients whose gold is already at rank 1 in the structural backbone suffer a $-13.7$~pp demotion cost ($p{=}5.8{\times}10^{-11}$): 35 patients move out of $R@1$ after reranking. This is the tax the reranker pays for being willing to move candidates at all, and it is an irreducible artifact of the Diagnostician's willingness to re-order the slate.

\emph{Rank 2--5 ($n{=}214$).} The load-bearing tier. FullMA lifts $R@1$ on this stratum by $+52.8$~pp ($p{=}1.9{\times}10^{-34}$); 113 of 214 patients move into $R@1$ after reranking. This stratum is the ``retrievable'' tier --- the gold disease was already close enough to the top that Gemma's mechanism-aware reasoning can discriminate it from the immediate neighbors.

\emph{Rank 6--10 ($n{=}152$).} Intermediate lift of $+31.6$~pp ($p{=}7.1{\times}10^{-15}$): 48 of 152 patients move into $R@1$. The lift is meaningful but attenuates with distance from the top; by rank 10 the rerank signal is competing with more noise.

\emph{Rank 11--30 ($n{=}161$).} The tier the 30-candidate slate unlocks: $+9.3$~pp $R@1$ ($p{=}6.1{\times}10^{-5}$), 15 patients moved to rank 1. At a 10-candidate slate this tier showed zero lift because Gemma never saw these candidates; widening the slate to 30 brings them into view, and the lift here (together with the $R@10$ expansion in \S\ref{sec:results:slate}) is what distinguishes the slate-30 headline from the slate-10 submission.

\emph{Rank past 30 ($n{=}340$, 30\% of cohort).} Unreachable. The gold disease is not in the gene-rescue pool at all, so no amount of reranking recovers it. This is the retrieval-coverage ceiling.

\textbf{Net flip.} Aggregating across strata, 176 patients gain $R@1$ (previously wrong, now correct) and 35 patients lose $R@1$ (previously correct, now wrong): a 5.0:1 favorable flip ratio. The fact that some correct top-1s are demoted is the honest cost of the rerank; stakeholder-facing messaging should include this rather than hide it.

\textbf{Framing implication.} The stratified table reframes the headline number. Rather than ``CausalOdyssey lifts $R@1$ by 12.57 percentage points,'' the accurate framing is ``on the 70\% of RareBench patients where retrieval is viable, CausalOdyssey concentrates its lift in the tier 2--5 band with $+52.8$~pp, extends into the rank 11--30 tier with $+9.3$~pp once the slate is wide enough to see it, at the cost of a $-13.7$~pp demotion tax on already-correct rank-1 patients.'' We argue that future LLM-retrieval work should report this stratification by default.

\subsection{Slate width: a recall--precision dial that saturates}\label{sec:results:slate}
The headline arms fix the Diagnostician slate width at 30 candidates. This width is itself a tunable: how many of the gene-rescue pool's candidates the LLM reranks. We ran the full $n{=}1{,}122$ pipeline at three widths (10, 30, 50) holding everything else fixed (Table~\ref{tab:slate}).

\begin{table}[h]\centering\small
\caption{Pooled $R@K$ vs.\ Diagnostician slate width, $n{=}1{,}122$, FullMA stack. $R@1$ peaks at width 30 ($+4.19$~pp over width 10, exact McNemar $p{=}4.3{\times}10^{-5}$) and is flat-to-down at 50; $R@5$/$R@10$ keep rising. Width 50 also widens the gene-rescue pool to 50.\label{tab:slate}}
\begin{tabular}{lccccc}
\toprule
slate & $R@1$ & $R@5$ & $R@10$ & $R@30$ & mean rank \\
\midrule
10 (submission) & 0.311 & 0.478 & 0.553 & 0.697 & 5.95 \\
30 & \textbf{0.353} & 0.513 & 0.593 & 0.697 & 4.97 \\
50 & 0.350 & \textbf{0.545} & \textbf{0.611} & \textbf{0.710} & 4.78 \\
\bottomrule
\end{tabular}
\end{table}

\textbf{$R@1$ saturates at width 30.} Widening from 10 to 30 lifts $R@1$ by $+4.19$~pp, concentrated in exactly the tiers the stratification flags as slate-starved: rank 6--10 and the newly-unlocked rank 11--30. Extending to width 50 does \emph{not} lift $R@1$ further ($0.350$ vs.\ $0.353$) but continues to lift $R@5$ ($+3.2$~pp) and $R@10$ ($+1.8$~pp): deep candidates that become visible at width 50 are promoted into the top-5/top-10, not to top-1.

\textbf{Tier-1 erosion is monotonic in width.} The cost of each added distractor is paid by the easy cases: rank-1 $R@1$ falls $0.878 \to 0.863 \to 0.851$ as the slate widens from 10 to 30 to 50. This erosion is the force that cancels the deep-band recall gain at $R@1$ past width 30, making the dial saturate. Width 30 is the $R@1$ optimum; width 50 is the $R@5$/$R@10$ optimum.

\textbf{Why we stop at 50.} An offline analysis of gold structural rank (to pool depth 200) shows recall saturating: $69\%$ of patients have gold within rank 30, $76\%$ within 50, $82\%$ within 100, $85\%$ within 200 --- the remaining $15\%$ are absent from gene rescue's reachable set and unrecoverable by any width. The marginal band beyond width 50 (66 patients at rank 51--100) is both small and the least convertible to $R@1$, while tier-1 erosion grows with every distractor. Widening the slate does not reach that cohort; moving gold up the structural ranking needs new information about the patient, such as a well-chosen follow-up finding (\S\ref{sec:closedloop}).

\subsection{Agent-removal reruns measure LLM rerun variance}\label{sec:results:ablation}
We re-ran the full pipeline on RAMEDIS $n{=}624$ with one downstream agent removed at a time, at slate widths 10 (reference FullMA $R@1 = 0.364$) and 30 (reference $0.436$). Table~\ref{tab:ablation} reports the result. The removed agents run after the Diagnostician and cannot change its ranking, so in a deterministic pipeline every entry would be exactly zero. They are not zero because each rerun calls the Diagnostician again at temperature 0.3: all 57 patients whose $R@1$ outcome differs between FullMA and FullMA$-$Verifier at slate-30 also have different Diagnostician scores. The table is therefore a direct measurement of single-run variance, not of agent contributions.

\begin{table}[h]\centering\small
\caption{$\Delta R@1$ between FullMA and a rerun with one downstream agent removed, RAMEDIS $n{=}624$ (exact paired McNemar $p$). The removed agents cannot change the ranking; the differences come from re-sampling the Diagnostician.\label{tab:ablation}}
\begin{tabular}{lcc}
\toprule
Remove & slate-10 $\Delta R@1$ ($p$) & slate-30 $\Delta R@1$ ($p$) \\
\midrule
Verifier     & $-1.28$~pp ($0.36$) & $-1.76$~pp ($0.19$) \\
Narrator     & $+2.40$~pp ($0.063$) & $+0.80$~pp ($0.58$) \\
Interrogator & $+0.80$~pp ($0.52$) & $+0.16$~pp ($1.00$) \\
\bottomrule
\end{tabular}
\end{table}

The spread is instructive. The largest entry ($+2.40$~pp, $p{=}0.063$) would read as ``the Narrator hurts retrieval'' if taken at face value. Single-run $R@1$ differences of $1$--$2.5$~pp on $n{=}624$ are within what re-sampling the same Diagnostician produces, and none of the six entries is significant. Any comparison between two single runs of an LLM reranker at temperature $>0$ should be read against this floor. The downstream agents are evaluated on what they do produce: the Verifier's verdict (\S\ref{sec:results:calibration}), the Interrogator's follow-up proposal (\S\ref{sec:closedloop}), and the Narrator's validated prose (\S\ref{sec:method:agents}).

\begin{figure}[h]
\centering
\includegraphics[width=0.82\linewidth]{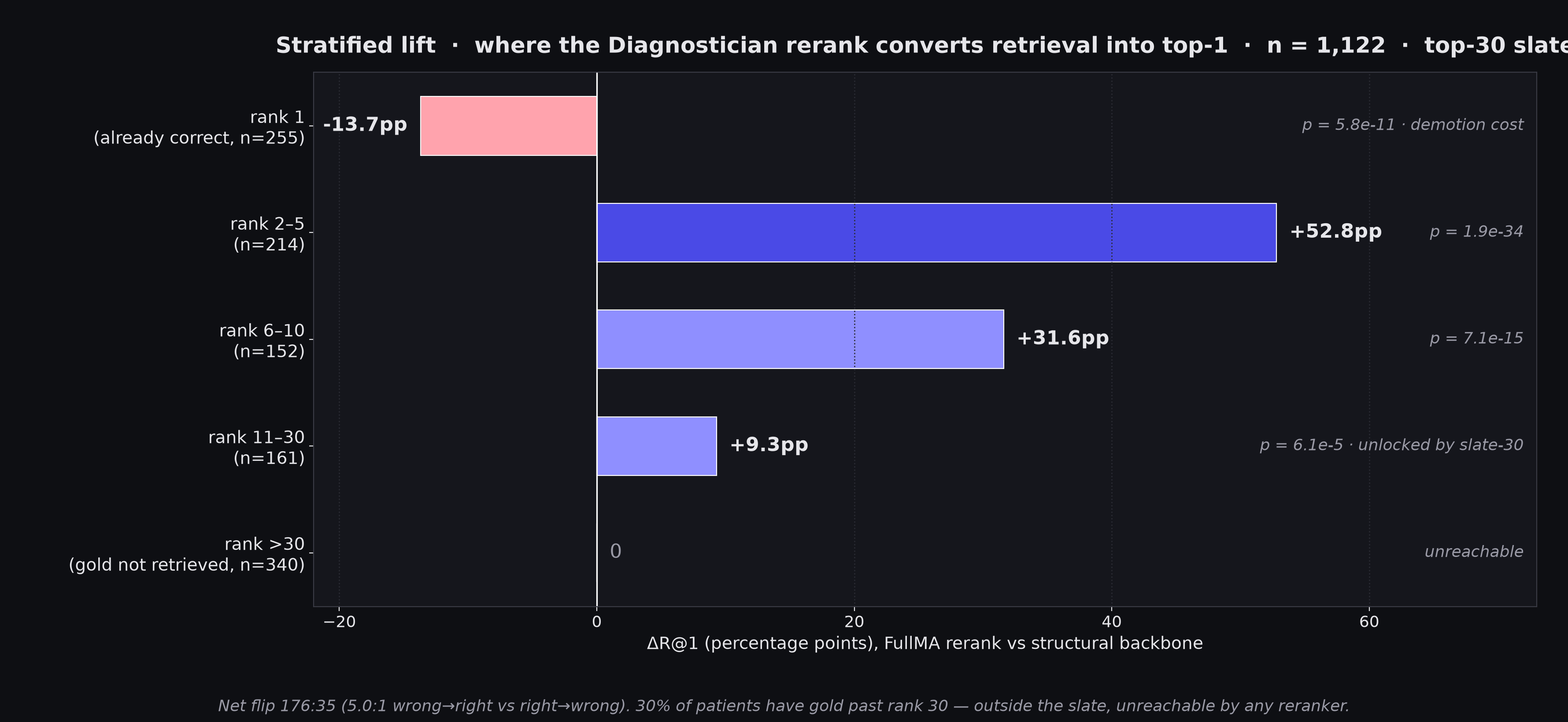}
\vspace{-0.5em}
\caption{\textbf{Stratified $\Delta R@1$ by baseline rank} (slate-30). The Diagnostician rerank concentrates lift in the retrievable tier (+52.8\,pp on rank 2--5, +9.3\,pp on rank 11--30); 30\% of patients with gold past rank 30 are unreachable. Rank-1 demotion cost is $-$13.7\,pp.}
\label{fig:strat}
\end{figure}

\subsection{Rejection-signal calibration}\label{sec:results:calibration}
We evaluate the Verifier's verdict as a rejection signal for the Diagnostician's top-1 prediction. On the $n{=}1{,}108$ patients in the slate-30 run with a valid Verifier verdict (for the other 14, the Verifier's Gemma call failed, its rationale failed citation validation, or there was no top-1), the verdict trichotomy distributes as \{confirm: $57.6\%$, flag: $29.8\%$, refute: $12.6\%\}$. The verdict is Gemma's judgment over the deterministic evidence (\S\ref{sec:method:agents}), so these numbers characterize that combination.

\textbf{Posterior drop.} The conditional accuracy given each verdict is: $\mathrm{P}(\text{top-1 correct}\,|\,\text{confirm}) = 51.4\%$; $\mathrm{P}(\text{top-1 correct}\,|\,\text{flag}) = 15.5\%$; $\mathrm{P}(\text{top-1 correct}\,|\,\text{refute}) = 12.1\%$. Relative to the prior $\mathrm{P}(\text{top-1 correct}) = 35.3\%$ (pooled $R@1$), a \emph{confirm} verdict raises the posterior to $1.5\times$ prior while a \emph{refute} verdict drops it to about a third of the prior and under a quarter of confirm --- a $4.2\times$ confirm-to-refute discrimination. A clinician receiving a \emph{refute}-flagged top-1 should substantially lower their prior that the pipeline's top pick is correct and expand their differential accordingly.

At slate-30 the three verdicts are monotonically ordered in correctness ($51.4\% > 15.5\% > 12.1\%$ for confirm/flag/refute), unlike the slate-10 run where \emph{flag} was anomalously the least-clean signal. \emph{Flag} and \emph{refute} remain close ($15.5\%$ vs $12.1\%$) and both sit far below \emph{confirm}; for clinical deployment we recommend treating \emph{flag} and \emph{refute} equivalently as ``expand the workup'' signals.

\textbf{Confidence-fusion AUC.} Treating the correctness of the top-1 as a binary classification task (is the ranker's top choice actually the gold?), we compare two fusion strategies on the slate-30 run. The Diagnostician confidence alone achieves AUC $= 0.726$; multiplying by the verdict multiplier (v2, \S\ref{sec:method:agents}) lifts AUC to $0.780$. An earlier fusion (v1, measured on the slate-10 run) also included a Narrator-validation rate term and a structural-anchor term; that version achieved AUC $= 0.605$, substantially worse, because the structural-anchor alone discriminates at AUC $= 0.514$ (near-random) and drags the signal. We ship v2 as the default.

\textbf{Is it the evidence or the judgment?} To test whether the deterministic evidence alone carries the rejection signal, we scored the same $n{=}1{,}108$ patients with rules fixed in advance that use only the code-computed evidence. A natural fixed rule (refute on a missing obligate phenotype or on failed closure with unexpected findings; flag on lesser findings) separates top-1 accuracy $52.7\%$ (confirm) from $27.5\%$ (refute), a $1.9\times$ spread against the Gemma verdict's $4.2\times$. Closure level alone is not monotone (tight $46.4\%$, loose $25.9\%$, none $35.6\%$). Gating Diagnostician confidence with any fixed rule lowers AUC (best $0.703$, vs.\ $0.726$ ungated and $0.782$ with the Gemma verdict). A cross-validated logistic model on Diagnostician confidence plus all evidence features reaches $0.748$, still below the verdict-gated score ($\Delta = -0.034$, 95\% bootstrap CI $[-0.063, -0.003]$); adding the verdict to that model gives $0.799$. The verdict therefore carries information beyond the evidence it is given: the rejection signal is an LLM judgment constrained to code-computed evidence, and that combination outperforms fixed rules over the same evidence.

\textbf{Narrator validation rate.} The $24.9\%$ citation-check pass rate (\S\ref{sec:method:agents}) reflects the strict validator rather than a weak Narrator. A looser validator would raise the pass rate at the cost of more hallucinated citations reaching the clinician.

\section{Does the Interrogator ask for the right test? A masked-phenotype evaluation}\label{sec:closedloop}

The Interrogator is meant to answer the clinician's next question: \emph{what should I test next?} Evaluating that offline is harder than it looks. A tempting design treats a proposal as a positive test result whenever HPOA annotates the proposed phenotype to the gold disease; but the Interrogator chooses phenotypes from those same annotations, and the structural backbone scores with them, so such an oracle rewards the system for agreeing with its own knowledge base. In the cohort below, that annotation oracle would call $17.3\%$ of the Interrogator's proposals positive, while only $0.6\%$ are phenotypes the patient actually has.

\subsection{Protocol}
We instead hide part of each patient's own record. For every RareBench patient with at least four observed HPOs, we mask two observed terms at random (five seeds), recompute the structural ranking on the remaining terms, and keep patients whose gold disease is then at rank 2--10 ($503$ patients, $1{,}501$ patient-seeds, all four corpora). The Interrogator proposes one phenotype, scoring candidates from the masked structural top-5 as in Algorithm~1. The simulated test is positive only if the proposal is one of the two hidden terms; a positive result is appended and the deterministic backbone re-run. The primary outcome is the \emph{hit rate}; we also report $\Delta R@1$ after a positive test. Comparators: a random annotation of the structural top-1, its most frequent annotation, and the highest-IC term in the top-5 pool, plus a ceiling that always asks for a hidden term. The protocol, cohort, and decision rule were written down before the masked runs. Unrecorded phenotypes are treated as absent, so every arm's hit rate is a lower bound.

\subsection{Results}
\begin{table}[h]\centering\small
\caption{Masked-phenotype follow-up, $1{,}501$ patient-seeds (two hidden terms per patient). 95\% patient-cluster bootstrap CIs. $\Delta R@1$: change in gold at rank 1 after the simulated follow-up.\label{tab:masked}}
\begin{tabular}{lcc}
\toprule
Proposal policy & Hit rate & $\Delta R@1$ (pp) \\
\midrule
Interrogator (Algorithm~1) & $0.6\%$ [0.1, 1.2] & $+0.20$ \\
Interrogator + mechanism boost & $0.7\%$ [0.2, 1.4] & $+0.13$ \\
Random annotation of top-1 & $1.2\%$ [0.6, 1.9] & $0.00$ \\
Most frequent annotation of top-1 & $1.2\%$ [0.5, 2.1] & $0.00$ \\
Highest-IC term in top-5 pool & $0.1\%$ [0.0, 0.2] & $+0.07$ \\
\midrule
Ceiling: ask for a hidden term & $100\%$ & $+11.13$ \\
\bottomrule
\end{tabular}
\end{table}

The Interrogator does not ask for the patient's hidden phenotypes more often than simple comparators (Table~\ref{tab:masked}; pre-specified exact McNemar on hits, seed 0: vs.\ random $p{=}0.69$, vs.\ most frequent $p{=}1.0$), and its proposals almost never move the gold diagnosis. Results are similar on RAMEDIS alone ($0.9\%$ vs.\ $1.0\%$ for random) and when the hidden terms are restricted to high-IC findings ($0.9\%$ vs.\ $2.4\%$ for most frequent), and counting ancestors of a hidden term as hits changes nothing material. The ceiling row shows that the question is worth answering: a follow-up that found one hidden finding would put the gold diagnosis at rank 1 for $11\%$ more patients.

\emph{Why it fails (exploratory).} A hidden term is often available to the Interrogator --- one is in the top-5 candidates' annotation pool for $61\%$ of patient-seeds --- but split-variance $\times$ IC ranks it low: it is the top proposal $0.6\%$ of the time and in the top 20 only $17\%$. Only $37\%$ of patients' recorded phenotypes are annotated to their own gold disease. Algorithm~1 picks the phenotype that best separates the candidates' \emph{annotations}; what patients actually present with is largely outside those annotations.

\emph{An illustrative case.} The live demo replays one case where a follow-up does help: for RAMEDIS\_179 (methylmalonic aciduria, cblB type), adding the proposed HP:0033443 (elevated propionylcarnitine, obligate in HPOA for the gold disease) moves the gold from position 6 to 1. It shows the intended interaction, not typical performance.

\subsection{What this means}
We do not claim that the Interrogator's proposals are informative. We contribute the evaluation: a masked-phenotype protocol that avoids the annotation-oracle circularity, a ceiling showing how much a good follow-up could add, and a negative result for an annotation-separating heuristic. Scoring candidate phenotypes by how often patients with each candidate disease actually present with them, rather than by how well they separate annotations, is the obvious next design to test under this protocol.

\section{Discussion}\label{sec:discussion}

\textbf{What the agents buy.} The retrieval lift is the Diagnostician's: it alone sets the ranking, and it moves pooled $R@1$ from 22.73\% to 35.29\% at slate-30. That lift is relative to our production baseline: a no-LLM rescoring matches it pooled, and the LLM rerank wins only on the metabolic corpus (\S\ref{sec:results}). On RAMEDIS, where pathway context discriminates closely related metabolic disorders, the reranker adds $16$~pp over even the stronger baseline; on case-report corpora it underperforms simple rescoring. The other three agents do not add $R@1$; they add outputs a single reranker does not produce. The Verifier gives a rejection signal ($4.2\times$ confirm-to-refute discrimination, and an AUC gain from $0.726$ to $0.780$ when its verdict gates the Diagnostician's confidence). The Interrogator gives a follow-up proposal, which on our masked-phenotype test does no better than simple baselines (\S\ref{sec:closedloop}). The Narrator gives citation-validated prose. The Verifier's discrimination comes from Gemma's judgment over the deterministic evidence, not from the evidence alone (\S\ref{sec:results:calibration}): fixed rules over the same evidence give at most a $1.9\times$ confirm-to-refute spread and lower AUC than Diagnostician confidence alone. What the design does guarantee is that the verdict is argued from evidence the model did not produce; whether a verdict model distinct from the Diagnostician would discriminate better is untested.

\textbf{Single runs are noisy.} Our agent-removal reruns changed $R@1$ by up to $2.4$~pp although the removed agents cannot affect the ranking (\S\ref{sec:results:ablation}). Read naively, that number looks like a finding about the Narrator. Multi-agent LLM papers routinely attribute single-run differences of this size to components; we recommend reporting the rerun spread of the unchanged pipeline alongside any ablation, and checking whether the removed component sits on the path that determines the metric at all.

\textbf{Retrieval-bounded lift, honestly reported.} Our stratified analysis (Figure~\ref{fig:strat}) shows that the pooled $+12.57$~pp concentrates in the retrievable tier: $+52.8$~pp on patients whose gold disease is already at rank 2--5 in the backbone, $+31.6$~pp on rank 6--10, $+9.3$~pp on rank 11--30, and $0$~pp on the 30\% of patients whose gold is past rank 30 (the gene-rescue pool depth). This decomposition matters because rerank-only lifts are frequently reported as single-number pooled averages, which creates a ceiling illusion: the reader assumes the method would continue to lift with more compute, when in fact the ceiling is set by upstream retrieval, and even the slate width that the reranker sees is a saturating dial (\S\ref{sec:results:slate}). We therefore frame our contribution as a \emph{rerank-within-retrievable} improvement, and argue that future work on LLM-augmented rare-disease retrieval should separately report the coverage-ceiling percentage and the rerank-among-covered lift. Moving the ceiling is a different problem (gene-rescue coverage, graph-embedding retrievers, free-text mining over case reports); doing more reasoning over a fixed retrieval slate will not solve it.

\textbf{Deployment implications for rural and low-connectivity settings.} All inference is local. The pipeline runs on one H100 in a median of $41$ seconds per patient ($14$--$50$~s for the middle $80\%$) (dominated by the Diagnostician's Gemma call; the other three agents' calls add $\approx 12$~s), with a cached-showcase mode that returns instantly on pre-computed patients for demonstration. Cloud-API dependencies are zero at runtime. This architecture is explicitly designed for settings like the (illustrative) Apur\'{\i}mac clinic in the introduction: the clinic does not need bandwidth for cloud inference, and the phenotype tuples the system reasons over never leave the premises. Our results use a single H100; Gemma~4~26B's mixture-of-experts architecture (4B active parameters) suggests that workstation-class local deployment is feasible, but we have not yet benchmarked consumer hardware, and we do not claim laptop-grade inference. Operationally, deployment in a rural setting would layer one additional component: a local HPO-extraction step for clinical prose, which we demonstrate via Gemma vision on a medical image but have not evaluated at scale. We believe the local-inference design is the most important practical contribution of the paper; the $+12.57$~pp retrieval lift is valuable, but what makes it deployable is that no cloud endpoint is involved.

\section{Limitations}\label{sec:limit}

\textbf{Research-grade, not clinical-grade.} CausalOdyssey is decision support, not decision. The chain-closure audit catches some but not all hallucinations --- it flags cases where Gemma's cited Reactome pathway fails to MONDO-bridge to the predicted disease, but it cannot detect fabrication of pathway IDs the audit has never seen, nor hallucinated mechanistic claims that happen to use a real pathway. A clinician must read the Verifier's verdict as a prior-shifting signal, not as ground truth. The system is not FDA-cleared and should not be deployed as a diagnostic tool without prospective validation.

\textbf{Retrieval-coverage ceiling.} 30\% of RareBench patients ($n{=}340$) have their gold diagnosis past rank 30 --- absent from the gene-rescue pool entirely. Since our reranker operates on the pool the backbone produces, no amount of LLM reasoning (and no slate widening, \S\ref{sec:results:slate}) can recover these cases: they are invisible to the reranker by construction. Our $+12.57$~pp pooled $R@1$ lift should be read as a concentrated gain on the 70\% of patients where retrieval is viable, not as a uniform lift across the cohort. A post-hoc audit of the 340 patients with gold rank past the top-30 cutoff is informative for where the next gains live: $99.9\%$ have their gold disease in our production database, and $83.5\%$ have nonzero overlap between patient HPOs and the gold gene's annotated phenotypes. On the high-overlap subcohort ($n{=}91$ with $\geq 6$ overlapping HPOs), the gold disease has strictly higher information-content-weighted overlap than the slate's top-1 in $94.5\%$ of cases --- suggesting, at first glance, that the gene-rescue scoring formula (the inverse-document-frequency penalty on multi-disease genes and the \texttt{n\_genes} normalization, which over-suppress canonical Mendelian genes such as PMM2, the gold gene for four patients in this cohort) is the bottleneck. We tested re-scoring directly with a structural-only re-implementation that reproduces the production ranks for 1{,}048 of 1{,}122 patients. Dropping the inverse disease-count weight lowers $R@1$ on the already-correct patients by $18.4$~pp but raises it overall, from $22.4\%$ to $36.3\%$ (and unreachable $R@30$ from $1.8\%$ to $30.9\%$); a square-root weight costs $3.1$~pp on the already-correct patients and reaches $28.6\%$ overall; cosine overlap and dropping both the weight and the normalization cost $42.3$ and $63.9$~pp on the already-correct patients. Without any LLM, the no-IDF variant matches the full pipeline's pooled $R@1$ ($36.3\%$ vs $35.3\%$, $p{=}0.58$) with a different profile: higher on LIRICAL ($53.0\%$ vs $27.0\%$) and MME ($57.5\%$ vs $27.5\%$), lower on RAMEDIS ($27.6\%$ vs $43.6\%$). On LIRICAL part of its gain is the annotation leakage discussed below, but it persists on the uncited cases ($35.7\%$ vs $15.2\%$ for the production baseline). Our production baseline is therefore not the strongest structural baseline, and we did not run the Diagnostician on the no-IDF candidates (\S\ref{sec:results}). These patients are not beyond reach of phenotype-based retrieval in general: LIRICAL places the gold diagnosis in its top-30 for 183 of the 340 (top-10 for 132, rank 1 for 35). The ceiling comes from our backbone's scoring and candidate list, not from missing data. Gated re-scoring (66 configurations: three gate signals $\times$ two variants $\times$ eleven thresholds), designed to protect the already-correct patients, found one configuration that cleared a no-regression guard ($+2.1$~pp unreachable $R@30$, firing on $12\%$ of patients) and missed the net gain above.

\textbf{Closed-loop reasoning is also retrieval-bounded.} A natural conjecture is that the unreachable cohort can be partially rescued by a closed-loop Interrogator: propose one follow-up HPO, append it to the patient profile, re-run gene-rescue, see whether gold rises into the top-30. We ran this experiment on the strict slate-30 unreachable sub-cohort (HPOA-verified follow-up at $\geq$Frequent, $n{=}50$, two pilots over the 30 cached candidates: a top-5 menu and a menu over all 30). Both pilots produced $0\%$ verified-rescue rate: the Interrogator's split-variance scorer cannot nominate a phenotype the gold actually has, because by construction the gold disease is not in the structural top-5 it discriminates over. Closed-loop reasoning is therefore not a lever for the unreachable cohort under our protocol; the same retrieval-bounded framing that limits LLM reranking also limits LLM-proposed follow-up. On the rerankable cohort the Interrogator also fails the masked-phenotype test (\S\ref{sec:closedloop}).

\textbf{Single-run results and confounded comparisons.} Every reranker number is one Gemma run at temperature 0.3. Paired bootstrap CIs capture patient-level variance but not run-to-run variance, which our reruns put at up to $\approx 2.5$~pp $R@1$ on RAMEDIS. The single-agent arm A3 was run only at slate-10 and with an earlier Diagnostician prompt, so we cannot attribute the A3-to-FullMA difference to any one change, and we have no slate-30 single-agent comparison. LIRICAL shows the smallest lift ($+2.97$~pp, $p{=}0.12$); self-consistency does not change that ($-0.27$~pp; Appendix~\ref{app:sc3}).

\textbf{HMS scope mismatch.} HMS ($n{=}88$, autoimmune/rheumatologic disease) has 51\% of patients' gold diagnoses past top-30 --- the gene-rescue retrieval backbone fundamentally does not fit this clinical domain, which lacks the clean one-gene-one-disease Mendelian mapping that drives metabolic and neurodevelopmental cases. Our lift on HMS ($+1.14$~pp $R@1$, CI [$-4.55$, $+6.82$], a net of one patient) reflects this ceiling. We retain HMS in the pooled number for completeness but scope the thesis to Mendelian rare disease in the headline framing; autoimmune reasoning is out of scope for this work.

\textbf{Annotation leakage from case reports.} HPOA curates disease annotations partly from published case reports, and several RareBench corpora are built from such reports. For LIRICAL we matched each of the 370 RareBench patients to its source case in the LIRICAL benchmark: for 199 of them (54\%), the HPOA annotations of the gold disease cite that case's own paper. On those patients the structural baseline is inflated (R@1 $31.7\%$, R@30 $79.9\%$, vs.\ $15.2\%$ and $55.0\%$ on the other 171), so absolute LIRICAL scores overstate performance for any HPOA-based method. The rerank lift does not come from this overlap: it is $+0.5$~pp on the cited patients and $+5.8$~pp on the uncited ones (12 gain vs 2 lose, $p{=}0.013$). We could not run the same check for RAMEDIS, HMS, or MME, whose source cases lack paper identifiers in RareBench; a leave-source-out rebuild of HPOA is needed to remove the effect.

\textbf{Masked phenotypes are a proxy for follow-up tests.} The masked-phenotype protocol (\S\ref{sec:closedloop}) avoids using HPOA as the oracle, but it only simulates follow-up on findings a clinician eventually recorded, and it treats unrecorded findings as absent, so hit rates are lower bounds. Real follow-up tests may target findings never recorded. Prospective evaluation with expert-adjudicated follow-ups is needed to measure clinical value.

\textbf{Gemma~4~26B specificity and language coverage.} All results use Gemma~4~26B (quantized) via Ollama. We have not evaluated smaller Gemma variants, non-Gemma models, or model-scale ablations; our code is model-agnostic but empirical results may not transfer. Similarly, the pipeline's front-end vignette extractor is demoed on English-plus-Spanish input (with a Quechua audio frame for motivation), but systematic multilingual evaluation of free-text $\to$ HPO extraction is future work.

\textbf{Hand-tuned design, and the case for an automated outer loop.} Most architectural choices in this paper --- the gene-rescue scoring formula, the missing-pathognomonic decay $\alpha$, the slate width, the closure-bonus weights --- were selected by manual search: propose a variant, evaluate $R@1$ on a held-out cohort, accept or reject against a regression guard on the already-correct tier. This is precisely the loop that the emerging self-evolving-agent literature proposes to automate. Meta-agent program search~\cite{adas} and self-modifying code agents that empirically validate their own edits against a benchmark~\cite{dgm}, together with self-adapting weight updates~\cite{seal}, offer a principled way to search the design space we explored by hand; recent surveys map the broader taxonomy~\cite{self_evolve_survey}. A natural next step is to let a meta-agent automate this search under our tier-1 regression guard and select on RareBench $R@1$ --- the guard is not incidental, since the same literature reports that open-ended self-improvement surfaces objective hacking (reward hacking)~\cite{dgm}, and an unconstrained search would likely rediscover the magnet-disease grab-bag failure the inverse-document-frequency penalty was introduced to suppress. We note, however, that the search space matters more than the search engine: our manual sweep of the scoring formula used a guard that only protected already-correct patients and so rejected a variant that improves the cohort overall (above). The objective an automated loop optimizes matters as much as the search itself. We also flag a methodological caution that bears directly on our own results: attributing end-to-end gains to the right component of an agent system --- base model, harness updates, or the agent's use of the harness --- is non-trivial~\cite{harness_updating}. Our own evaluation shows the evaluation-side version of this problem: comparing our slate-10 single-agent run with the slate-30 pipeline run (which also used a revised Diagnostician prompt), or reading the agent-removal reruns as per-agent effects (\S\ref{sec:results:ablation}), would credit gains to the wrong components. We recommend that future self-evolving retrieval systems hold the harness fixed and report rerun variance when attributing gains to evolution.

\textbf{Data and ethics.} All patient data used in this work come from public, de-identified benchmark corpora (RareBench and its constituent datasets, HPOA, MONDO, Reactome, Orphanet); no new human-subjects data were collected. The introductory clinic vignette is an illustrative composite, not a real case.

\section{Availability}\label{sec:availability}
Demo code (MIT): \url{https://github.com/xueerchen1990/CausalOdyssey}. The public repository holds the demo version (May 2026); the evaluation code that produced the results reported here differs from it, notably in the Verifier's Gemma verdict step and the Interrogator's scoring function, and will be released in the same repository.\ \ \ \textbullet\ \ \ Live demo (HuggingFace Space): \url{https://huggingface.co/spaces/xmiao/causal-odyssey}\ \ \ \textbullet\ \ \ Model: Gemma~4~26B via Ollama (Apache 2.0).\ \ \ \textbullet\ \ \ Data: RareBench, HPOA, OMIM/ORPHA, Reactome, MONDO --- all public.

\bibliographystyle{unsrt}
\bibliography{causalodyssey}


\appendix

\section{Prompt templates}\label{app:prompts}
All four agents call Gemma with structured prompts. We reproduce the Diagnostician prompt used in the evaluation, the Narrator system prompt, and the Verifier system prompt here. The full set (mechanism-aware variant, few-shot exemplars, cluster-aware variant, Interrogator rationale prompt) is in \texttt{src/gemma\_rerank.py}, \texttt{src/agents/narrator.py}, \texttt{src/agents/gemma\_verifier.py}, and \texttt{src/agents/interrogator.py}.

\subsection{Diagnostician rerank prompt}\label{app:prompts:rerank}
The evaluation used the mechanism-aware rerank template in \texttt{src/gemma\_rerank.py} (with this template the chain-of-thought flag has no effect). Placeholders are filled with two retrieved worked examples, the routed HPOs, up to ten bidirected FCI pairs, and the 30 candidates with their genes, sample phenotypes, and Reactome blocks. Long lines are wrapped here.

\begin{lstlisting}
You are assisting with rare disease diagnosis. Score how well each candidate disease
  mechanistically explains the patient.

{fewshot}

=== PATIENT OBSERVED PHENOTYPES ===
{observed_phenotypes}

=== SHARED-MECHANISM PHENOTYPE PAIRS (iGFCI bidirected edges) ===
{bidirected_edges}

=== CANDIDATE DISEASES (with Reactome pathway evidence where available) ===
Each candidate lists its supporting gene(s), its canonical phenotypes, AND Reactome
  pathways. [D]=disease-specific pathway; [B]=biology-level pathway. PMIDs are from each
  pathway's literature reference list.

{candidates}

=== TASK ===
Score EACH candidate 0-100 on mechanistic fit:
- Does the candidate's gene-pathway explain the patient's observed phenotypes,
  especially rare/specific ones?
- Does the candidate have a DISEASE-SPECIFIC ([D]) Reactome pathway matching the patient
  presentation? That is strong mechanism evidence.
- Biology-level ([B]) pathways without a disease tag count only as weak evidence; score
  on HPO fit when that's all there is.

SCORING RULES:
- Use the FULL 0-100 range. Distinguish candidates from each other.
- At most TWO candidates may score 85+. Spread the rest (e.g., 92, 78, 65, 55, 45).
- Use INTEGER scores; don't default to 50 or 100 when uncertain.

REASONING STRUCTURE: for each candidate's `reason`, follow this order in under 25 words:
  1. Name 1-2 observed phenotypes the candidate's gene-pathway best explains.
  2. Note any distinguishing feature present (raises score) or absent (lowers score).
  3. State the score rationale in a half-clause.
Example: "Orotic aciduria + hyperammonemia fit OTC urea-cycle block; R-HSA-9956553; no
  male-sib history lowers prior."

CITATION RULE (grounding):
- Every R-HSA id or PMID you cite for a candidate MUST appear in that same candidate's
  Reactome evidence block above.
- If no valid citation exists, omit pathway_refs and pmid_refs -- do not invent, do not
  borrow from other candidates.

Output ONLY a JSON object:
{"scores": [
  {"disease_id": "OMIM:xxx", "score": <int 0-100>, "reason": "<reasoning-structure, may
    cite R-HSA-* and PMIDs from this candidate's block>", "pathway_refs": ["R-HSA-xxx"],
    "pmid_refs": ["xxxxx"]}
]}
Omit pathway_refs / pmid_refs when none. No prose or markdown outside the JSON.
\end{lstlisting}

\subsection{Narrator system prompt}\label{app:prompts:narrator}
The Narrator runs with a strict system prompt that enumerates allowed and forbidden citation sources. The user prompt supplies an \texttt{<evidence\_block/>} assembled from the trace.

\begin{lstlisting}
You are a clinical summarizer producing a short, auditable diagnosis
narrative. You follow strict rules.

ALLOWED:
- Cite disease IDs (OMIM:X / ORPHA:X) and disease names only from the
  <evidence_block/>
- Cite HPO IDs (HP:nnnnnnn) only if they appear in the <evidence_block/>
- Cite gene symbols only if they appear in the <evidence_block/>
- Cite Reactome pathway IDs (R-HSA-nnnnnnn) only if they appear in the
  <evidence_block/>
- Cite PubMed IDs (PMID:n) only if they appear in the <evidence_block/>
- Reference chain closure level by its label ("tight", "loose", "none",
  "unclosable") when the Verifier reports one
- Describe mechanism at textbook level using concepts already named in
  the evidence

FORBIDDEN:
- DOIs, author names, journal names, URLs
- Disease names or IDs not in the evidence block
- Gene symbols not in the evidence block
- HPO IDs not in the evidence block
- Reactome pathway IDs not in the evidence block
- PMIDs not in the evidence block
- Probabilistic claims ("80% of patients...") beyond the frequency labels
  in the evidence
- Speculation about unseen phenotypes unless the evidence flags them as
  expected-but-missing

You will return structured JSON matching the schema. Every claim must map
to a trace key path listed in cited_trace_keys. Prefer sentences that cite
a pathway (R-HSA-*) or PMID (PMID:*) when such evidence is present - these
are the most auditable citations.
\end{lstlisting}

The Narrator user prompt (\texttt{NARRATOR\_PROMPT\_TEMPLATE}) supplies the evidence block and asks for a JSON object with \texttt{top\_diagnosis\_sentence}, \texttt{supporting\_evidence\_sentences} (2--4 sentences, each citing one trace key), \texttt{caveats\_sentence}, \texttt{next\_action\_sentence}, \texttt{cited\_trace\_keys}, \texttt{confidence}, \texttt{abstained}, and \texttt{abstain\_reason}. The validator (\texttt{src/agents/validators.py}) enforces that every cited entity resolves to an entry in the trace; unresolved citations are stripped before the narrative is surfaced.

\subsection{Verifier system prompt}\label{app:prompts:verifier}
The Verifier's deterministic stage (\texttt{src/agents/verifier.py}) computes the three checks of \S\ref{sec:method:agents}; the Gemma stage (\texttt{src/agents/gemma\_verifier.py}) receives them in the user prompt under this system prompt:

\begin{lstlisting}
You audit a rare-disease diagnosis. You do NOT re-rank.

Your job: given the top-1 candidate disease, the patient's phenotypes, and
a deterministic pathognomonic-missing check, decide whether the top-1 is
biologically plausible.

Return ONLY a JSON object:
{"verdict": "confirm" | "flag" | "refute",
 "rationale": "<1-3 sentences, under 200 chars>",
 "confidence_adjustment": -0.2 to +0.1,
 "cited_hpos": ["HP:...", ...]}

Rules:
- "confirm" = top-1 fits the patient's pattern well; biology consistent.
- "flag" = partial fit, ambiguous; keep top-1 but mark low confidence.
- "refute" = strong biological mismatch (e.g. obligate phenotype
  definitively absent, wrong mechanism). Use rarely.
- Cite HPO IDs ONLY if they appear in the PATIENT_OBSERVED or
  EXPECTED_MISSING lists in the evidence.
- NO disease IDs in rationale unless they appear in the evidence block.
- NO literature citations, DOIs, author names.
[... one rule on confidence_adjustment (-0.2 to +0.1) and the
 rationale-required block omitted ...]
\end{lstlisting}

The prompt continues with instructions that the rationale be non-empty (abridged here). The Interrogator's HPO is chosen by Algorithm~1 in \texttt{src/agents/interrogator.py} against the HPOA frequency table, patient FCI edges, and the Reactome membership index; a separate Gemma call (\texttt{build\_gemma\_interrogator\_rationale}) writes a rationale for the chosen HPO.

\section{Pydantic schemas}\label{app:schemas}
All agent I/O is validated against strict Pydantic v2 schemas. Parse failures fall through to a deterministic stub path (\S\ref{sec:method:safety}). Full definitions are in \texttt{src/agents/schemas.py}; key fields are reproduced below.

\textbf{DiagnosticianOutput.} The reranked slate (up to 30 candidates) with reasoning summary, per-candidate confidence, and an abstention hatch.

\begin{lstlisting}
class DiagnosticianOutput(BaseModel):
    ranked_candidates: List[RankedCandidate]
    reasoning_summary: str = ""
    confidence: float = Field(0.0, ge=0.0, le=1.0)
    abstained: bool = False
    abstain_reason: Optional[str] = None

class RankedCandidate(BaseModel):
    disease_id: str = Field(pattern=DISEASE_PATTERN)
    disease_name: str
    confidence: float = Field(ge=0.0, le=1.0)
    mechanism_fit_sentence: str
    supporting_hpos: List[str] = []
    supporting_genes: List[str] = []
    fci_edge_refs: List[str] = []
    pathway_refs: List[str] = []       # R-HSA-* pathway IDs
    pmid_refs: List[str] = []          # PMID:* references
    chain_closure: Optional[Literal["tight", "loose", "none"]] = None
\end{lstlisting}

\textbf{VerifierOutput.} Trichotomy verdict on top-1, missing and unexpected HPOs, chain-closure audit dict, and confidence adjustment.

\begin{lstlisting}
class VerifierOutput(BaseModel):
    top1_disease_id: Optional[str] = None
    expected_but_missing: List[MissingPhenotype] = []
    unexpected_present: List[MissingPhenotype] = []
    confidence_adjustment: float = Field(0.0, ge=-1.0, le=1.0)
    flagged_low_confidence: bool = False
    rationale: str = ""
    verdict: Optional[str] = None   # confirm / flag / refute (Gemma)
    chain_closure_audit: Optional[Dict[str, Any]] = None
    confidence: float = Field(0.0, ge=0.0, le=1.0)
    abstained: bool = False
    abstain_reason: Optional[str] = None

class MissingPhenotype(BaseModel):
    hpo: str = Field(pattern=r"^HP:[0-9]{7}$")
    name: str
    ic: Optional[float] = None
    frequency: Optional[str] = None   # e.g., "Obligate (100%)"
\end{lstlisting}

\textbf{InterrogatorOutput.} Proposed next HPO with its expected discrimination gain and an abstention hatch.

\begin{lstlisting}
class InterrogatorOutput(BaseModel):
    proposed_hpo: Optional[str] = None
    proposed_name: Optional[str] = None
    rationale: str = ""
    expected_in_diseases: List[str] = []
    unexpected_in_diseases: List[str] = []
    expected_discrimination_gain: float = 0.0
    confidence: float = Field(0.0, ge=0.0, le=1.0)
    abstained: bool = False
    abstain_reason: Optional[str] = None
\end{lstlisting}

\textbf{NarratorOutput.} Citation-validated clinical prose with trace-key references.

\begin{lstlisting}
class NarratorOutput(BaseModel):
    top_diagnosis_sentence: str
    supporting_evidence_sentences: List[str]
    caveats_sentence: Optional[str] = None
    next_action_sentence: Optional[str] = None
    cited_trace_keys: List[str]
    confidence: float = Field(ge=0.0, le=1.0)
    abstained: bool = False
    abstain_reason: Optional[str] = None
\end{lstlisting}

\section{Reproducibility}\label{app:reproducibility}
\textbf{Model.} Gemma~4~26B~\cite{gemma4} (the 26B-A4B mixture-of-experts variant, quantized; 25.2B total parameters, 3.8B active per token; open weights under Apache~2.0) served via Ollama~\cite{ollama} (\texttt{ollama pull gemma4:26b}; the same tag is hard-coded as the default in the public repository's Diagnostician and Narrator, so the eval and the released code resolve to the same library entry).
We have not evaluated other quantization levels or non-Gemma backbones.

\textbf{Inference parameters.} All Gemma calls use $N{=}1$ sample and \texttt{think=false}; temperature and token limits are per agent (\S\ref{sec:eval}). The \texttt{think=false} flag is critical: without it, Gemma~4 routes reasoning tokens to the \texttt{.thinking} field of the Ollama response and \texttt{.response} returns empty.

\textbf{Determinism.} The structural backbone is fully deterministic (deterministic ordering on tied gene-rescue scores); evaluation bootstrap resamples use a seeded NumPy RNG. Gemma inference is not deterministic at temperature 0.3; reruns produce slightly different top-5 orderings on boundary cases. For the headline table we report a single-run number; paired bootstrap CIs capture the patient-level variance but not the Gemma-run variance, which the agent-removal reruns (\S\ref{sec:results:ablation}) put at up to $\approx 2.5$~pp $R@1$ on RAMEDIS.

\textbf{Wall-clock.} Structural backbone: $<500$~ms per patient. Measured in the slate-30 run (medians, interquartile range): Diagnostician 34.5~s (9.2--35.4); Verifier Gemma call 1.6~s (1.5--1.8); Interrogator rationale call 1.1~s (1.1--1.3); Narrator 9.5~s (3.2--11.3). Full pipeline: median 41~s per patient (10th--90th percentile 14--50~s) on one NVIDIA H100.

{\sloppy \textbf{Data versions.} HPO and HPOA release \texttt{2026-02-16} (\texttt{hp.obo}, \texttt{phenotype.hpoa}, \texttt{phenotype\_to\_genes.txt}); MONDO rare-disease subset release \texttt{2025-06-03}; Reactome API queried April 2026; RareBench unified benchmark~\cite{rarebench}, as released 2024. The demo's \texttt{scripts/download\_data.sh} fetches the HPOA and MONDO files at clone time; all other data is bundled in the repository or the cached showcase JSONs.\par}

\section{Self-consistency ablation (SC3)}\label{app:sc3}
LIRICAL's small and unstable lift (\S\ref{sec:results}) motivated an ablation with $N{=}3$ Diagnostician samples per patient (self-consistency over three independent Gemma calls at temperature 0.3), with the final top-5 chosen by majority vote on the top-1 candidate.

\textbf{Results (slate-30).} Relative to FullMA ($N{=}1$) at the headline 30-candidate slate, SC3 changes $R@1$ by: RAMEDIS $-1.76$~pp ($0.436\to0.418$, $n{=}624$); LIRICAL $-0.27$~pp ($0.270\to0.268$, $n{=}370$); HMS $+1.14$~pp ($n{=}88$); MME $+15.00$~pp ($0.275\to0.425$, $n{=}40$). Pooled $\Delta R@1$ is approximately flat ($34.85\%$ vs $35.29\%$, $-0.45$~pp). Two findings differ from what we observed at a 10-candidate slate, where SC3 had appeared to recover LIRICAL (+4.32~pp): at slate-30 the LIRICAL effect is negligible ($-0.27$~pp), and the apparent recovery was therefore largely a slate-10 artifact. The only sizeable positive is MME, but at $n{=}40$ ($\approx 6$ patients) this is within noise. Self-consistency at slate-30 is best read as a near-neutral-to-slightly-negative intervention on this benchmark, not the per-dataset tradeoff the slate-10 run suggested.

\textbf{Implication.} We do not ship SC3: at the headline slate it provides no reliable pooled benefit and a measurable RAMEDIS cost. The earlier hypothesis that $N{\ge}3$ self-consistency cleanly recovers LIRICAL does not survive the slate correction; the slate-10 difference was within single-run variance (\S\ref{sec:results:ablation}).

\end{document}